\documentclass[useAMS,usenatbib]{mn2e}
\usepackage[a4paper,margin=2cm]{geometry}
\usepackage{times}
\usepackage{epsfig}
\usepackage[]{graphicx} 
\usepackage{pdfpages}
\usepackage{verbatim} 
\usepackage[fleqn]{amsmath}
\usepackage{amssymb}
\usepackage{calc}
\usepackage{caption}
\usepackage[normal]{threeparttable}
\usepackage{longtable}
\usepackage{pdflscape}
\newcommand{\comments}[1]{} 
\usepackage{placeins} 
\usepackage{float}

\title[Associated H\,{\normalsize \it I} absorption in HIPASS]{A
  search for H\,{\Large\bf I} absorption in nearby radio galaxies
  using HIPASS} \author[J.~R. Allison et
al.]{J.~R. Allison$^{1,2}$\thanks{E-mail: jra@physics.usyd.edu.au},
  E.~M. Sadler$^{1,2}$ and A.~M. Meekin$^{1}$\\$^{1}$Sydney Institute
  for Astronomy, School of Physics A28, University of Sydney, NSW
  2006, Australia\\$^{2}$ARC Centre of Excellence for All-sky
  Astrophysics (CAASTRO)}
\begin{document}

\date{}

\pagerange{\pageref{firstpage}--\pageref{lastpage}} \pubyear{2012}

\maketitle

\label{firstpage}

\begin{abstract} Using archival data from the \mbox{H\,{\sc i}} Parkes
  All Sky Survey (HIPASS) we have searched for 21\,cm line absorption
  in 204 nearby radio and star-forming galaxies with continuum flux
  densities greater than $S_{1.4} \approx 250$\,mJy within the
  redshift range $0 < cz < 12\,000\,\mathrm{km}\,\mathrm{s}^{-1}$. By
  applying a detection method based on Bayesian model comparison, we
  successfully detect and model absorption against the radio-loud
  nuclei of four galaxies, of which the Seyfert~2 galaxy
  2MASX\,J130804201-2422581 was previously unknown. All four
  detections were achieved against compact radio sources, which
  include three active galactic nuclei (AGNs) and a nuclear starburst,
  exhibiting high dust and molecular gas content.  Our results are
  consistent with the detection rate achieved by the recent ALFALFA
  (Arecibo Legacy Fast Arecibo L-band Feed Array) \mbox{H\,{\sc i}}
  absorption pilot survey by Darling et al. and we predict that the
  full ALFALFA survey should yield more than three to four times as
  many detections as we have achieved here. Furthermore, we predict
  that future all-sky surveys on the Square Kilometre Array precursor
  telescopes will be able to detect such strong absorption systems
  associated with type 2 AGNs at much higher redshifts, providing
  potential targets for detection of H$_{2}$O megamaser emission at
  cosmological redshifts.  \end{abstract}


\section{Introduction}\label{section:introduction} Atomic hydrogen
(\mbox{H\,{\sc i}}) gas, traced by the 21\,cm line, is a powerful
probe of the mass distribution within galaxies and the available fuel
for future star formation. However, the strength of the 21\,cm
emission-line decreases rapidly with increasing redshift as a function
of the inverse square of the luminosity distance. In individual
galaxies the most distant detectable 21\,cm emission lines are at
$z\sim0.2$ \citep[e.g.][]{Catinella:2008, Verheijen:2010,
  Freudling:2011}, while statistical detections using spectral
stacking have reached $z\approx0.37$ \citep{Lah:2009}. At
significantly higher redshifts, we can instead detect the 21\,cm line
in individual galaxies through the absorption of continuum flux
towards a background radio source. In principle, \mbox{H\,{\sc i}}
absorption can be observed up to cosmological redshifts, where the
ionosphere begins to corrupt the signal, yet such observations are
ultimately limited by the sample of known high-redshift radio sources
and the availability of suitable instrumentation. The highest
redshifts achieved by observations of 21\,cm absorbers include the
radio galaxy B2\,0902+34 ($z = 3.397$; \citealt{Uson:1991}) and the
intervening system towards the quasar PKS\,0201+113 ($z = 3.387$;
\citealt{Kanekar:2007}). The detection limit for a survey of 21\,cm
line absorption is independent of redshift, and depends only on the
availability of bright background continuum sources against which the
line can be detected. For any such sight line, absorption is
particularly sensitive to high column densities ($N_\mathrm{HI}
\gtrsim 10^{20}\,\mathrm{cm}^{-2}$) of cold ($T_\mathrm{spin} \lesssim
100$\,K) foreground \mbox{H\,{\sc i}} gas that obscures a large
fraction of the background radio source.

In the local Universe, at least 10\,\,per\,cent of extragalactic radio
sources that have been searched exhibit an associated 21\,cm
absorption line at or near the optical redshift, indicating that
neutral gas is present within the host galaxy
\citep[e.g.][]{Morganti:2001, Vermeulen:2003, Allison:2012a}. High
signal-to-noise ratio (S/N) absorption lines, typically associated
with powerful radio galaxies, often exhibit broad wings, which can
indicate the presence of fast jet-driven outflows of \mbox{H\,{\sc i}}
gas (with velocities over 1000 km\,s$^{-1}$ and outflow rates of
several tens of M$_\odot$\,yr$^{-1}$; e.g.
\citealt{Morganti:2005b,Mahony:2013,Morganti:2013}). These
high-velocity outflows may have a profound effect on the star
formation and subsequent evolution of the host galaxy. Furthermore,
there is evidence to suggest that in some cases broad absorption
components can arise in circumnuclear gas distributed as a disc or
torus \citep[e.g.][]{Struve:2010a, Morganti:2011}. Such observations
are incredibly useful for directly studying the interaction between
the radio-loud nucleus and the neutral gas in the interstellar medium.
However, at present these surveys are limited to targeted sampling of
the radio source population, typically focusing on those that are
compact, and by doing so can introduce biases \citep[see
e.g.][]{Curran:2010}.

Here we present the results of a search for 21\,cm absorption in
nearby radio and star-forming galaxies from the \mbox{H\,{\sc i}}
Parkes All-Sky Survey (HIPASS; \citealt{Barnes:2001}). While the data
are considerably less sensitive than current targeted observations of
radio sources (with an effective integration time of 7.5\,min per
individual pointing and typical rms noise of
$\sim$13\,mJy\,beam$^{-1}$ per 13\,km\,s$^{-1}$ channel separation),
the large volume covered by HIPASS (the whole sky south of $\delta =
+25^\circ$ and $z\lesssim0.042$) allows identification of the
strongest associated \mbox{H\,{\sc i}} absorption-line systems in the
local universe in an unbiased way. This enables us to study some of
the most extreme and potentially interesting systems, as well as
testing line-finding techniques \citep[e.g.][]{Allison:2012b} that can
be used in planning future, more sensitive, large-area surveys with
the Square Kilometre Array (SKA) pathfinder and precursor telescopes.

\cite{Darling:2011} recently published the results of a pilot survey
for \mbox{H\,{\sc i}} 21\,cm absorption in the Arecibo Legacy Fast
Arecibo L-band Feed Array (ALFALFA) survey.  This was the first
genuinely blind search for absorption within a large-area radio
survey, and covered 517\,deg$^2$ of sky in the redshift range
$z<0.058$. No intervening lines were seen, but one previously known
associated line was re-detected in the interacting luminous infrared
galaxy UGC\,6081 \citep{Bothun:1983, Williams:1983}. The HIPASS search
presented here can be considered complementary to that survey, since
it covers a much larger area of sky (by approximately a factor of 50)
with similar redshift coverage, but has lower sensitivity (by
approximately a factor of 6). However, due to the presence of strong
baseline ripple, we have limited our search to the detection of
\mbox{H\,{\sc i}} absorption within the host galaxies of the radio
sources themselves. Spectral baseline ripples are a common problem for
single-dish observations of the 21\,cm line
\citep[e.g.][]{Briggs:1997} and HIPASS spectra towards bright
continuum sources are particularly affected, where standing waves are
generated between the primary dish and receiver cabin
\citep{Barnes:2001,Barnes:2005}. By using the known systemic redshift
of the galaxy as a prior, we can attempt to distinguish the
absorption-line from the strong baseline ripple. We intend in future
work to revisit the HIPASS data with improved analysis and perform an
extended search of intervening \mbox{H\,{\sc i}} absorption within the
full volume.

Throughout this paper we adopt a flat $\Lambda$ cold dark matter
cosmology with $H_{0}$ = 70\,km\,s$^{-1}$, $\Omega_\mathrm{M}$ = 0.3
and $\Omega_{\Lambda}$ = 0.7. Radial velocities and redshifts have
been corrected for the solar barycentric standard-of-rest frame.

\section{Sample selection}

Our sample selection was driven by the brightest radio sources in the
National Radio Astronomy Observatory Very Large Array Sky Survey
\citep[NVSS, $\nu = 1.4$\,GHz;][]{Condon:1998}, the Sydney University
Molonglo SkySurvey \citep[SUMSS, $\nu = 843$\,MHz;][]{Mauch:2003} and
the second epoch Molonglo Galactic Plane Survey \citep[MGPS-2, $\nu =
843$\,MHz;][]{Murphy:2007}. Together, the footprints of these three
surveys fully overlap the sky coverage of HIPASS down to continuum
flux densities of a few mJy. The typical noise per median-gridded
HIPASS image is $\sim$ 13\,mJy\,beam$^{-1}$ (with spectral channels
separated by 13.2\,km\,s$^{-1}$ at z = 0) but can vary significantly
as a function of system temperature and the number of gridded
pointings contributing to the image. By considering those radio
sources that have integrated flux densities above 250\,mJy (at either
843\,MHz or 1.4\,GHz), which would enable us to detect absorption
lines with peak optical depths greater than 30\,per\,cent against the
weakest sources, we have constructed a sample of 19\,237 within the
HIPASS footprint of $-90\degr < \delta < +25\degr$. To obtain a sample
of nearby radio and star-forming galaxies, we simply matched this list
of radio sources with their optical counterparts and selected those
that have redshifts within the HIPASS volume. However, to
significantly improve the completeness of our sample we also
considered the catalogue of \cite{VanVelzen:2012}, who have used a
more sophisticated method to match radio sources with their
counterparts in the Two Micron All-Sky Survey
\citep[2MASS][]{Skrutskie:2006} Redshift Survey \citep{Huchra:2012}.

\subsection{Sample 1: radio--optical matches}

In the first instance, we construct a sample of nearby radio and
star-forming galaxies that have known redshifts in the range $cz <
12\,000$\,km\,s$^{-1}$, by matching our catalogue of 19\,237 radio
sources with their optical counterparts using the \textsc{MultiCone}
search function of the \textsc{TopCat} software package
\citep{Taylor:2005}. The optical counterparts were selected using
catalogues from the 6dF Galaxy Survey \citep[6dF\,GS;][]{Jones:2009}
and the CfA Redshift Survey (\citealt{Huchra:1999} and references
therein), or otherwise from the NASA Extragalactic
Database\footnote{http://ned.ipac.caltech.edu/}. Based on the work of
\citet{Mauch:2007}, who matched radio sources in NVSS with galaxies in
6dF\,GS (for $0.003 < z < 0.3$), we assume that a maximum displacement
of 10\,arcsec is sufficient to produce a reliable identification of a
radio-optical source pair. Of these radio-optical pairs, we identified
105 with optical spectroscopic redshifts in the range spanned by the
HIPASS data. A further 15 matches were then excluded from the sample,
in most cases due to unreliable redshift measurements (see
Appendix\,\ref{section:sample1_exclusions}), resulting in a final list
of 90 nearby galaxies that form our first sample.

\subsection{Sample 2: the van Velzen et al. sample}

The recently compiled catalogue of nearby radio and star-forming
galaxies by \citet{VanVelzen:2012} was constructed by matching radio
sources in the NVSS and SUMSS catalogues with their optical
counterparts in the 2MASS Redshift Survey (2MRS;
\citealt{Huchra:2012}), covering 88\,per\,cent of the sky at redshifts
of $z \lesssim 0.052$. This catalogue consists of 575 galaxies with
apparent $K_{s}$-band magnitudes brighter than 11.75 and total flux
densities above limits\footnote{Based on the total 1.4\,GHz flux
  density of Centaurus\,A \citep[$S_{1.4} = 1330$\,Jy;][]{Cooper:1965}
  and assuming a spectral index of $\alpha = -0.6$, where $\log(S)
  \propto \alpha\log(\nu)$.} of 213\,mJy at 1.4\,GHz and 289\,mJy at
843\,MHz. Importantly, matches were made between multiple radio
components and a single galaxy, thereby providing reliable estimates
of the total radio flux density in extended emission. Given that the
MGPS-2 catalogue only identifies compact radio components, van Velzen
et al. did not consider these galactic-plane sources and so any sample
constructed from their catalogue will not contain sources with
Galactic latitude $|b| < 10\degr$ south of $\delta = -30\degr$. From
this parent catalogue, we have selected a second sample of 189
galaxies that are bounded by the volume $cz < 12\,000$\,km\,s$^{-1}$
and $-90\degr < \delta < +25\degr$.

\subsection{Properties of our sample}

By comparing the content of our two samples, we found that 75 of the
galaxies in Sample 1 are common to those in Sample 2, while the
remaining 15 either have radio flux densities or 2MASS $K_{s}$-band
magnitudes below the limits imposed by \cite{VanVelzen:2012}, or are
in the MGPS-2 compact source catalogue. Therefore, our total sample
contains 204 unique radio-detected galaxies. van Velzen et al. defined
morphological classifications for their radio galaxy sample based on
the extent and distribution of the radio emission compared with the
near-infrared emission. By applying those same classifications to our
sample, we find that 39 are \emph{point sources}, 124 are \emph{jets
  and lobes}, 36 are \emph{star-forming galaxies} and 5 are
\emph{unknown}.  Those galaxies that have an \emph{unknown}
classification are potentially the result of a random match with a
background radio source; however, we have decided to include these in
our sample since they still provide reasonable candidates for
\mbox{H\,{\sc i}} absorption.

\begin{figure}
\centering
\includegraphics[width = 1.0\columnwidth]{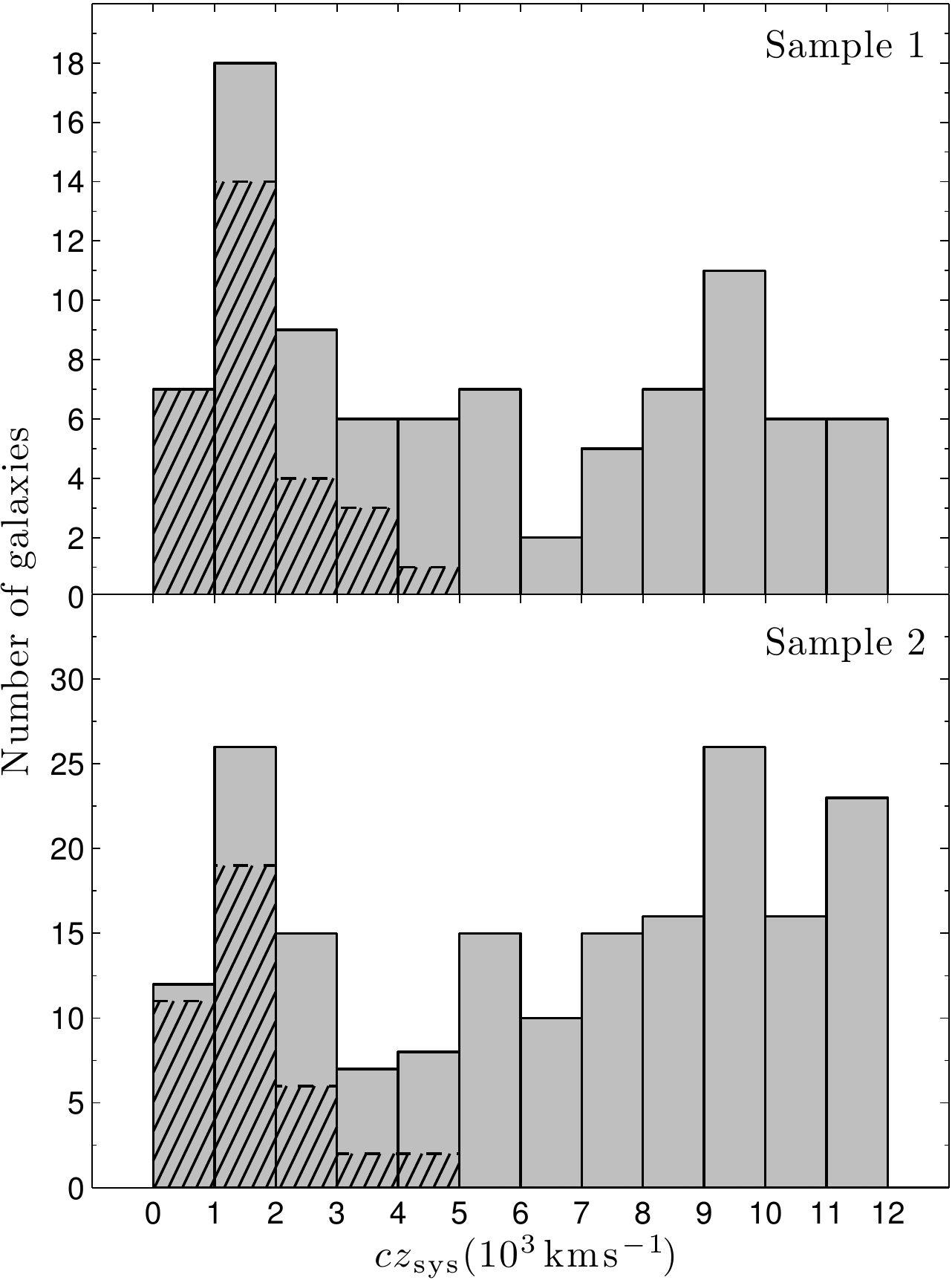}
\caption{The distribution of systemic redshifts for our two samples of
  galaxies. The hatched bars represent those with catalogued 21\,cm
  line emission in HIPASS (\citealt{Koribalski:2004, Meyer:2004,
    Wong:2006}), dominating at redshifts less than $cz_\mathrm{sys} =
  5000$\,km\,s$^{-1}$.}\label{figure:redshift_hist}
\end{figure}

The completeness of our combined sample will be limited by that of the
parent catalogues (for example the MGPS-2 compact source catalogue of
the galactic plane excludes the 10\,per\,cent of extended radio
sources predicted by SUMSS), and the matching algorithms employed (for
example our 10\,arcsec position-matching criterion for Sample 1 will
exclude some nearby large radio and star-forming galaxies). To provide
an estimate of the completeness, we compare the total number of radio
and star-forming galaxies in our sample with that predicted from the
local luminosity function at 1.4\,GHz. The HIPASS footprint covers an
area of sky equal to 29\,343\,deg$^{2}$ \citep{Meyer:2004,Wong:2006},
and so with an upper redshift limit of $cz < 12\,000$\,km\,s$^{-1}$
our sample spans a comoving volume of 0.0146\,Gpc$^{3}$. Based on the
local radio luminosity function given by \cite{Mauch:2007}, which was
measured from a sample of 6667 galaxies at $0.003< z < 0.3$, we
predict that there are approximately 230 galaxies within the HIPASS
volume above a flux density limit of 250\,mJy and 260 above
213\,mJy. Our sample of 204 galaxies therefore represents an
approximately 80--90\,per\,cent complete flux-limited list of nearby
radio and star-forming galaxies in the HIPASS footprint, not
accounting for the uncertainties generated by counting statistics,
cosmic variance and the effects of galaxy clustering. We list in
Appendix\,\ref{section:full_sample} the properties of candidates that
form our sample, and in Fig.\,\ref{figure:redshift_hist} we show their
distribution as a function of redshift.

\section{H\,{\sevensize\bf I} absorption in HIPASS}

\subsection{Spectra extraction}\label{section:spectra_extraction}

Calibration and imaging of the Parkes 21\,cm multibeam data was
described in extensive detail by \cite{Barnes:2001}, with further
descriptions of the final HIPASS emission-line catalogues by
\cite{Koribalski:2004}, \cite{Meyer:2004} and \cite{Wong:2006}. For
each galaxy in our two samples, we have searched for \mbox{H\,{\sc i}}
absorption in a single integrated spectrum towards the centroid
position of the radio source. The spectra were extracted from the data
cubes by implementing the task \textsc{Mbspect} in the Multichannel
Image Reconstruction, Image Analysis and Display
package\footnote{http://www.atnf.csiro.au/computing/software/miriad}
(\textsc{Miriad}; \citealt{Sault:1995}). The gridded beamwidth of each
HIPASS image is 15.5\,arcmin, and so we assume that any \mbox{H\,{\sc
    i}} absorption in the target galaxy will be detected in a single
pencil beam towards the spatially unresolved radio emission. The
overlaid radio contours and optical images shown in
Fig.\,\ref{figure:image_overlays_total} for our sample show that this
assumption is valid.

The flux density spectrum of each unresolved radio source is
calculated by taking the following weighted sum over a square region
of nine by nine 4\,arcmin pixels, as follows
\begin{equation} \label{equation:flux_measurement}
  S(\nu) = \sum_{i}{w_{i}^{\prime}}S_{i}(\nu),
\end{equation}
and
\begin{equation} \label{equation:weights} 
w_{i}^{\prime} = {w_{i} \over \sum_{i}{{w_{i}}^2}},
\end{equation}
where $S_{i}(\nu)$ and $w_{i}$ are the flux density and weighting,
respectively, for the $i$th pixel. For an elliptical beam, with
position angle $\phi$ and axes with full width at half maxima (FWHMs)
of $\theta_\mathrm{maj}$ and $\theta_\mathrm{min}$, the beam weights
are given by
\begin{equation} \label{equation:beam_weights} 
w_{i} = \exp{\left[-2\ln{(4)}(A_{i} + B_{i} + C_{i})\right]},
\end{equation}
where 
\begin{eqnarray}
  A_{i} & = & [(\sin{(\phi)}/\theta_\mathrm{maj})^{2}+(\cos{(\phi)}/\theta_\mathrm{min})^{2} ]~(\Delta{\alpha_{i}})^{2}, \nonumber\\
  B_{i} & = & \sin{(2\phi)}~[(1/\theta_\mathrm{maj})^{2} - (1/\theta_\mathrm{min})^{2}]~\Delta{\alpha_{i}}~\Delta{\delta_{i}}, \nonumber \\
  C_{i} & = & [(\cos{(\phi)}/\theta_\mathrm{maj})^{2}+(\sin{(\phi)}/\theta_\mathrm{min})^{2}]~(\Delta{\delta_{i}})^{2},
\end{eqnarray}
and $\Delta{\alpha_{i}}$ and $\Delta{\delta_{i}}$ are the angular
distances from the centre position, in right ascension and
declination, respectively. For the median-gridded HIPASS images, $\phi
= 0\degr$ and $\theta_\mathrm{maj} = \theta_\mathrm{min} =
15.5$\,arcmin, so in this case the beam weighting parameters are given
by
\begin{eqnarray}
  A_{i,\mathrm{HIPASS}} & = & (\Delta{\alpha_{i}}/15.5\mathrm{\,arcmin})^{2}, \nonumber \\
  B_{i,\mathrm{HIPASS}} & = & 0, \nonumber \\
  C_{i,\mathrm{HIPASS}} & = & (\Delta{\delta_{i}}/15.5\mathrm{\,arcmin})^{2}.
\end{eqnarray}
The extracted HIPASS spectra for each of our 204 galaxies are shown in
Fig.\,\ref{figure:spectra_total}.

\subsection{Noise estimation}

\begin{figure}
\centering
\includegraphics[width = 1.0\columnwidth]{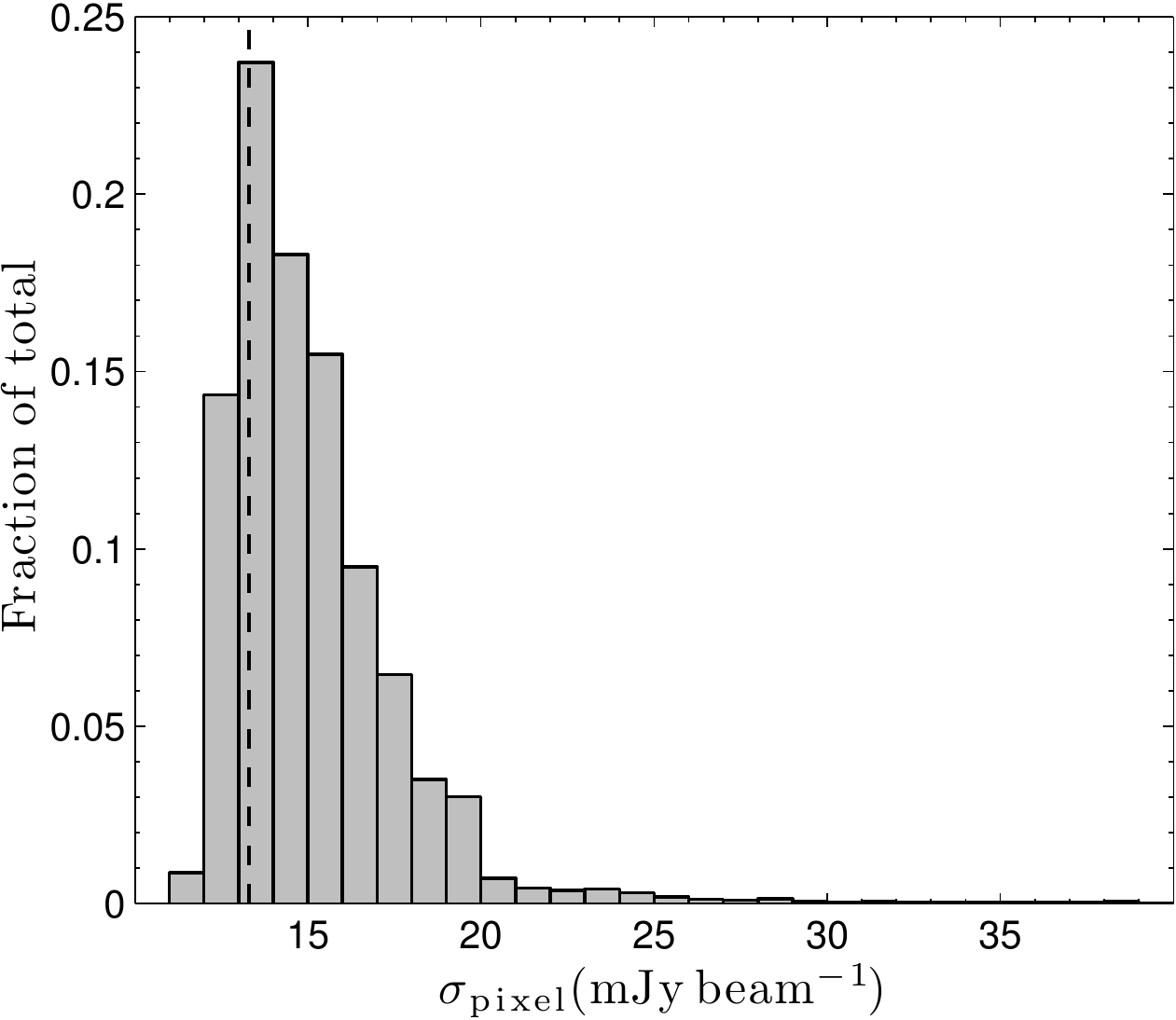}
\caption{The distribution of pixel noise (standard deviation) per
  HIPASS channel per sight-line for our sample of 204 galaxies,
  estimated using the MADFM over each image plane of each data
  cube. For comparison we show the canonical rms noise (vertical
  dashed line) for the HIPASS southern data (13.3\,mJy\,beam$^{-1}$;
  \citealt{Barnes:2001}).}\label{figure:hipass_rms}
\end{figure}

In order to estimate the significance of individual spectral
components of a given HIPASS spectrum, we must characterize the
properties of the noise. The archival HIPASS data cubes were
constructed by gridding together individual spectra using a median
estimator to the beam-weighted average \citep{Barnes:2001}. The median
of a randomly distributed variable is asymptotically normal, and so
given that the noise in the individual ungridded spectra is
approximately normal and that the beam weights are randomly
distributed on the sky, we assume that the noise in the median-gridded
spectra is also normal. In a given data cube, we allow for spectral
variation in the noise by calculating the median absolute deviation
from the median (MADFM) across each image plane. The MADFM per pixel
($s_{\rm pixel}$) is given by
\begin{equation}
s_{\rm pixel} = \mathrm{med}(|d_{i} - \mathrm{med}(d_{i})|),
\end{equation}
where $d_{i}$ is the value of the $i$th pixel in the image. The
standard deviation per pixel ($\sigma_{\rm pixel}$), assuming that the
pixel noise is normally distributed, can then be estimated by
(\citealt{Whiting:2012})
\begin{equation}
  \sigma_{\rm pixel} = [\sqrt{2}\mathrm{erf}^{-1}(0.5)]^{-1}\,s_{\rm pixel} \approx 1.483\,s_{\rm pixel},
\end{equation}
where $\mathrm{erf}^{-1}$ is the inverse of the Gauss error function.
Fig.\,\ref{figure:hipass_rms} shows the distribution of the estimated
pixel noise per channel per sight-line for our sample of galaxies,
peaking in the range 13--14\,mJy\,beam$^{-1}$.  This method provides a
robust estimator of the pixel noise for a given channel, but does not
account for spatial variation across the data cube. Furthermore, the
MADFM is a robust estimator of the noise for data where sources occupy
a relatively small number of pixels with respect to the total size of
the image, yet it may become a poor estimator in channels containing
extended strong signal, such as that from the 21\,cm line in the Milky
Way.

The standard deviation ($\sigma_{\rm S}$) in the flux density ($S$)
defined by Equation\,\ref{equation:flux_measurement} is given by
\begin{equation}
  \sigma_{\rm S}^{2} = \bmath{w}^{\rm t}\,\mathbfss{C}_{\rm pixel}\,\bmath{w},
\end{equation}
where $\bmath{w}$ is the vector of weights ($w_{\rm i}^{\prime}$)
defined by Equation\,\ref{equation:weights} and $\mathbfss{C}_{\rm
  pixel}$ is the image pixel covariance matrix. If we assume that the
per-pixel noise has a single value ($\sigma_{\rm pixel}$), we can
simplify this expression to
\begin{equation}
  \sigma_{\rm S}^{2} = (\bmath{w}^{\rm t}\,\mathbfss{R}_{\rm pixel}\,\bmath{w})\,\sigma_{\rm pixel}^{2},
\end{equation}
where $\mathbfss{R}_{\rm pixel}$ is the pixel correlation matrix. The
noise correlation between pixels is generated by a combination of the
intrinsic properties of the telescope (such as the beam) and the
gridding procedure implemented by Barnes et al. Rather than
analytically modelling this relationship between $\sigma_{\rm S}$ and
$\sigma_{\rm pixel}$, which would require knowledge of the relative
contributions of these factors, we estimate it empirically by
generating multiple Monte Carlo realizations of the flux density $S$
per image per data cube. Following this empirical procedure, we find
that
\begin{equation}
  \sigma_{\rm S} \approx (0.95 \pm 0.10)\,\sigma_{\rm pixel},
\end{equation}
and so apply a correction factor of 0.95 to the pixel noise when
estimating the noise level in our extracted spectra.

\begin{figure}
\centering
\includegraphics[width = 1.0\columnwidth]{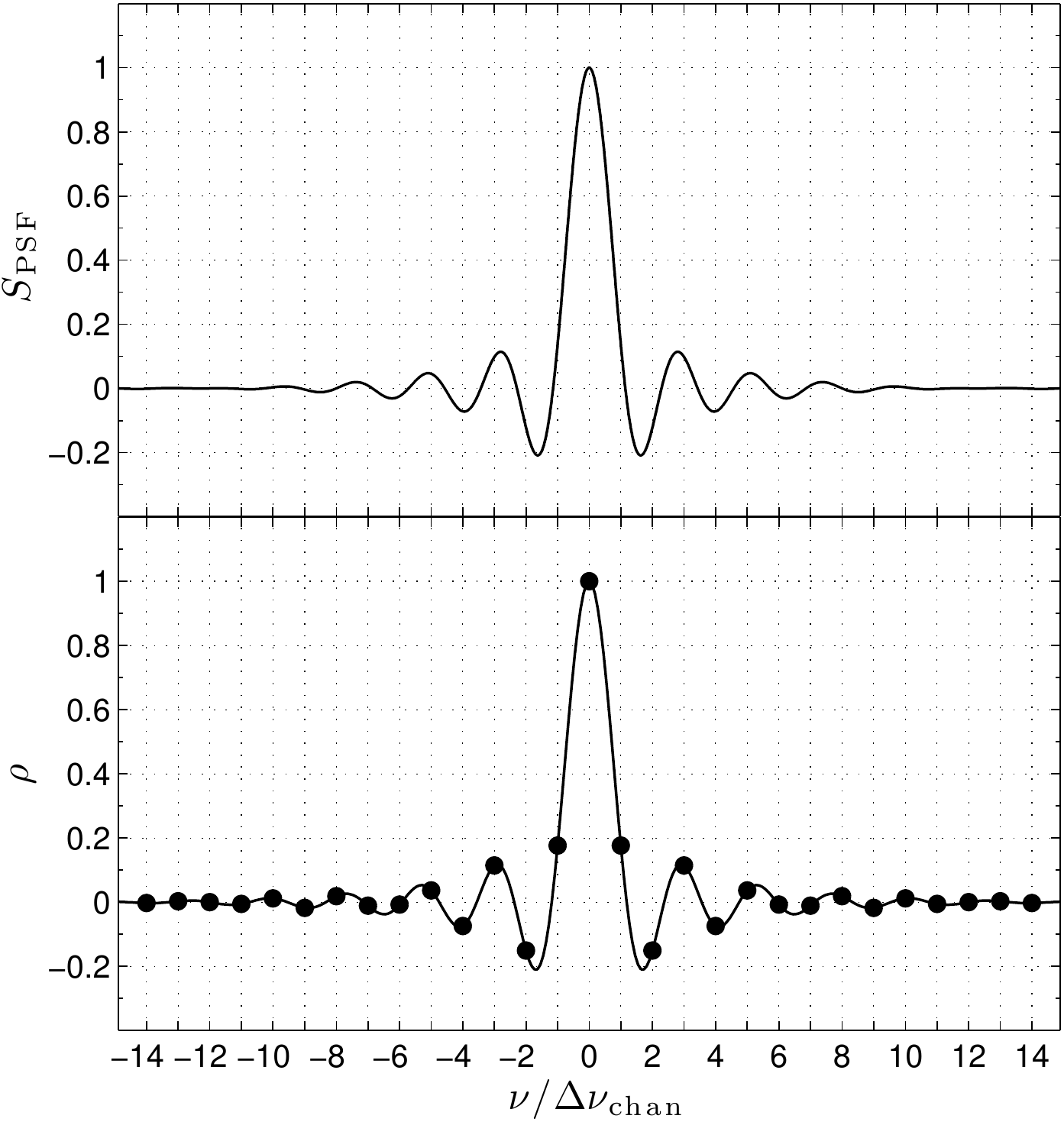}
\caption{Top: the spectral response of HIPASS to an unresolved
  positive signal centred at channel 0 and normalized to
  unity. Bottom: the resulting correlation function ($\rho$) given by
  Equation\,\ref{equation:correlation}. The circles represent the
  correlation coefficients sampled at the channel separation
  $\Delta{\nu_\mathrm{chan}}$.}\label{figure:hipass_correlation}
\end{figure}

\subsection{Covariance estimation}

A common consideration for lag correlators, such as that used for
HIPASS, is the effect of a strong signal in the unfiltered time-lag
spectrum, which introduces severe Gibbs ringing in the frequency
domain. To significantly reduce this effect, Barnes et al. applied a
25\,per\,cent Tukey filter to the time-lag data, diminishing the
spectral resolution by 15\,per\,cent and effectively increasing the
FWHM of the spectral point spread function to approximately
18\,km\,s$^{-1}$. Using this information, we can model the correlation
($\rho$) as a function of frequency ($\nu$) by taking the
autocorrelation of the known spectral point spread function
($S_\mathrm{PSF}$, top panel of
Fig.\,\ref{figure:hipass_correlation}), which in algebraic form is
given by
\begin{equation}\label{equation:correlation}
  \rho(\nu) = \rho_{0}^{-1}\int_{-\infty}^{\infty}{S_\mathrm{PSF}(\nu^{\prime})S_\mathrm{PSF}(\nu+\nu^{\prime})}\mathrm{d}\nu^{\prime},
\end{equation}
where 
\begin{equation}
  \rho_{0} = \int_{-\infty}^{\infty}{S_\mathrm{PSF}(\nu^{\prime})}^{2}\mathrm{d}\nu^{\prime}.
\end{equation}
The correlation coefficients between discrete HIPASS spectral channels
can then be calculated by sampling $\rho(\nu)$ at integer channel
separations (see the bottom panel of
Fig.\,\ref{figure:hipass_correlation}). Using these correlation
coefficients we can estimate the noise covariance ($\sigma_{ij}^{2}$)
between the $i$th and $j$th channels by
\begin{equation}
  \sigma_{ij}^{2} = \rho_{ij}\,\sigma_{i}\,\sigma_{j}, 
\end{equation}
where $\rho_{ij}$ is the correlation coefficient and $\sigma_{i}$ is
an estimate of the standard deviation due to the noise in channel
$i$. These covariances form the off-diagonal elements of the matrix
$\textbfss{C}$, while the per-channel variances ($\sigma_{i}^2$) form
the diagonal. We note that it is also possible, with enough
information, to model other components of the covariance. For example,
we could account for the aforementioned standing waves that dominate
single-dish spectra towards radio continuum sources or, in the case of
interferometers, the spectral ripple that can arise as a result of
imaging a field of continuum sources with incomplete sampling of the
Fourier plane. While this will be pursued in future work, here we
choose to consider the effects of these systematics a posteriori, and
therefore compare our spectral models based purely on their
significance above our estimate of the correlated noise.

\subsection{Automated line finding and
  parametrization}\label{section:line_finding}

We automatically detect and parametrize the \mbox{H\,{\sc i}}
absorption by using a Bayesian approach to model comparison, the
application of which was described by
\citet{Allison:2012a,Allison:2012b}. This method determines the
significance of a detection above the noise by comparing the posterior
probability of the absorption-line and continuum model
($\mathcal{M}_\mathrm{HI}$) with that of the continuum-only model
($\mathcal{M}_\mathrm{cont}$), given the data. Using Bayes' theorem,
the posterior probabilities of the two models are related to the
marginal likelihoods (also known as the evidence),
$\mathrm{Pr}(\bmath{d}|\mathcal{M})$, and priors,
$\mathrm{Pr}(\mathcal{M})$, by
\begin{equation}
  {\mathrm{Pr}(\mathcal{M}_\mathrm{HI}|\bmath{d})\over\mathrm{Pr}(\mathcal{M}_\mathrm{cont}|\bmath{d})} = {\mathrm{Pr}(\bmath{d}|\mathcal{M}_\mathrm{HI})\over\mathrm{Pr}(\bmath{d}|\mathcal{M}_\mathrm{cont})}~{\mathrm{Pr}(\mathcal{M}_\mathrm{HI})\over\mathrm{Pr}(\mathcal{M}_\mathrm{cont})},
\end{equation}
where $\bmath{d}$ is the data. By assuming that we are suitably
uninformed about the presence of an absorption line (so that the above
ratio of priors is unity), we define our detection statistic ($R$) by
\begin{equation}\label{equation:detection_statistic}
  R \equiv \ln\left[{\mathrm{Pr}(\bmath{d}|\mathcal{M}_\mathrm{HI})\over\mathrm{Pr}(\bmath{d}|\mathcal{M}_\mathrm{cont})}\right].
\end{equation}
We can estimate the marginal likelihood of the data for each model by
integrating the likelihood as a function of the model parameters
($\bmath{\theta}$), over the parameter prior,
\begin{equation}
  \mathrm{Pr}(\bmath{d}|\mathcal{M}) = \int{\rmn{Pr}(\bmath{d}|\bmath{\theta},\mathcal{M})~\rmn{Pr}(\bmath{\theta}|\mathcal{M})~\rmn{d}\bmath{\theta}},
\end{equation}
which is implemented using the Monte Carlo sampling algorithm,
\textsc{MultiNest} (developed by \citealt{Feroz:2008} and
\citealt{Feroz:2009b}). An efficient method for estimating the
uncertainty in this integral, and hence in our detection statistic $R$
is described by \cite{Skilling:2004} and \cite{Feroz:2008} and
implemented in \textsc{MultiNest}. The dominant uncertainty arises
from the statistical approach of nested sampling to estimating the
widths between likelihood samples contributing to this integral. This
decreases as the square root of the number of active samples used in
the algorithm, and increases as the square root of the information
content of the likelihood relative to the prior (the negative relative
entropy). Therefore, for a fixed number of active samples, the
absolute uncertainty in $R$ increases with both the S/N in the data
and the number of model parameters. For the analysis presented here,
we find that an active sample size of larger than 500 is sufficient to
provide uncertainties in $R$ that are smaller than unity (equal to a
relative probability of approximately 3 between the two marginal
likelihoods), while still maintaining computational efficiency.

Assuming that the data are well approximated by a normal distribution,
the likelihood as a function of the data and model is given in its
general form by
\begin{equation}\label{equation:likelihood}
  \rmn{Pr}(\bmath{d}|\bmath{\theta},\mathcal{M}) 
  =  {1\over\sqrt{(2\mathrm{\pi})^{N}|\mathbfss{C}|}}\exp{\left[-{(\bmath{d} - \bmath{m})^\rmn{t}\mathbfss{C}^{-1}(\bmath{d} - \bmath{m})\over2}\right]},
\end{equation}
where $\bmath{m}$ is the expected data given the model parameters, $N$
is equal to the total number of data, and $\textbfss{C}$ is the
covariance matrix. The model data $\bmath{m}$ are generated by
convolving our parametrization of the physical signal with the
spectral response function $S_\mathrm{PSF}$ shown in
Fig.\,\ref{figure:hipass_correlation}. We parametrize the 21\,cm
absorption line by the summation of multiple Gaussian components, the
best-fitting number of which can be determined by optimizing the
statistic $R$. Since \citet{Barnes:2001} reported that the spectral
baseline has been adequately subtracted, we assume that for the HIPASS
data the continuum component is best represented by the zero-signal
($\bmath{m} = 0$) model. For data where the continuum is still
present, this can be modelled using a simple polynomial representation
(see e.g. \citealt{Allison:2012a} and
Section\,\ref{section:cabb_analysis}).

\subsection{Model parameter priors}

For each model parameter, we use an informed prior based on the known
observational and physical limits. The following is a description of
the priors chosen for each of the absorption-line parameters.

\subsubsection{Redshift}

Since we are searching for \mbox{H\,{\sc i}} absorption associated
with the host galaxy of each radio continuum source, we can use
existing measurements of the systemic redshift to strongly constrain
the allowed redshift of each spectral line component. To this end, we
choose a normal prior with a mean value equal to the systemic redshift
(as given in Table\,\ref{table:full_sample}) and a 1\,$\sigma$ width
equal to 50\,km\,s$^{-1}$. Such a prior is consistent with the
uncertainties given for existing all-sky redshift surveys, e.g. 2MRS
(\citealt{Huchra:2012}), 6dFGS (\citealt{Jones:2009}) and the Sloan
Digital Sky Survey (SDSS; \citealt{Aihara:2011}), as well as the
typical differences in redshifts between these surveys (see
e.g. fig.\,5 of \citealt{Huchra:2012}). By using a sufficiently
constrained prior on the redshift, we can attempt to differentiate an
absorption line from the strong systematic baseline ripples known to
exist in the HIPASS spectra and therefore avoid excessive false
detections that would occur in a blind survey of
redshift-space. However, we do acknowledge that this could potentially
exclude those absorption-lines that arise in \mbox{H\,{\sc i}} gas
that is either rapidly in falling or outflowing with respect to the
active galactic nucleus (AGN). Furthermore, we note that while the
majority of galaxies in our sample have systemic redshift
uncertainties smaller than 50\,km\,s$^{-1}$, in a few cases some are
larger.

\begin{figure*}
\centering
\includegraphics[width = 1.0\textwidth]{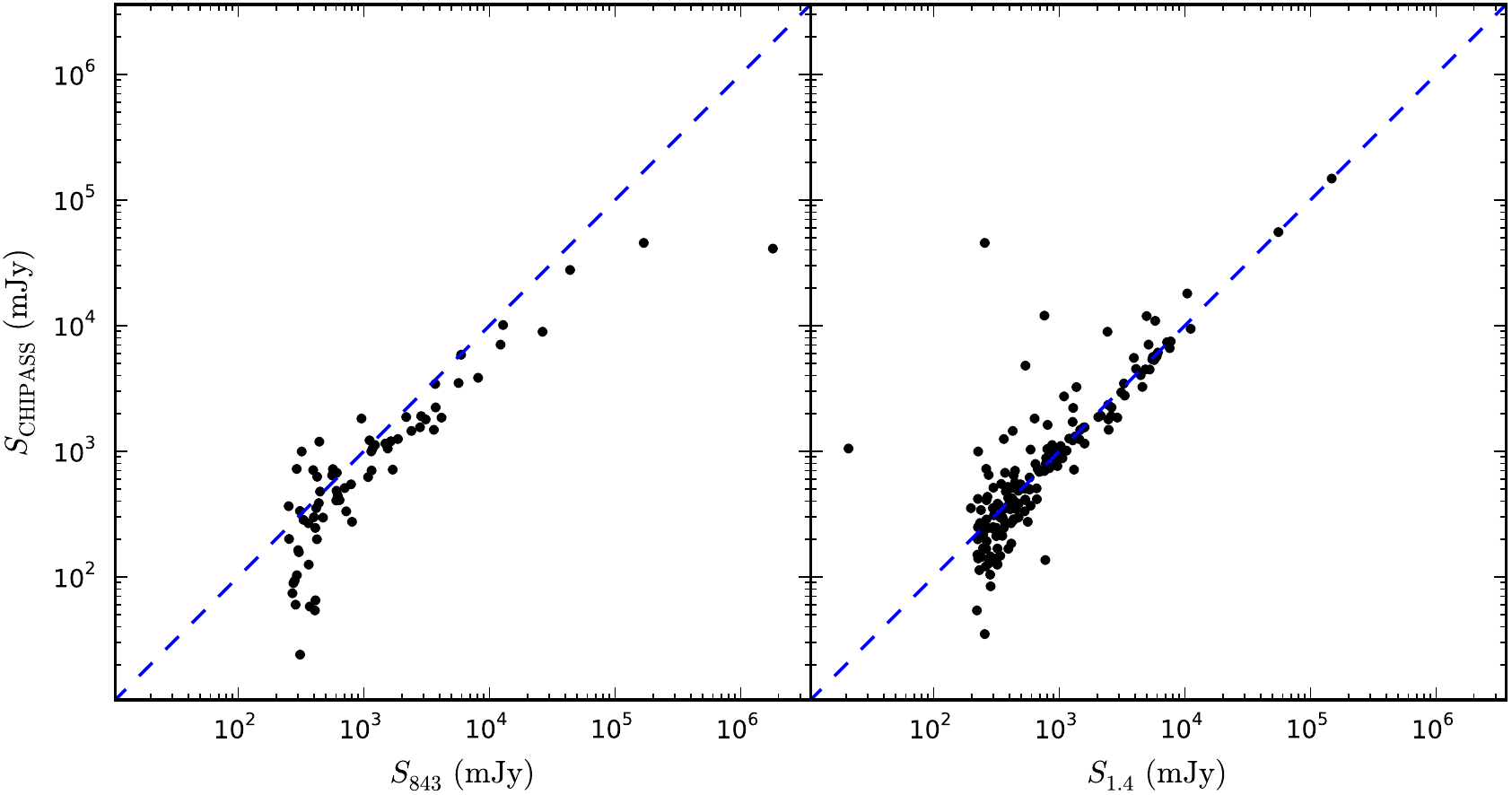}
\caption{The CHIPASS beam-weighted flux density versus the
  SUMSS/MGPS-2 (left) and NVSS (right) total flux density for 204
  galaxies in our sample. $y(x) = x$ is indicated by the dashed blue
  line.}
\label{figure:chipass_flux_comparison}
\end{figure*}

\subsubsection{Velocity width}

We assign a uniform prior to the line FWHM in the velocity range
0.1--2000\,km\,s$^{-1}$. Since the spectral channel separation and
resolution of the HIPASS data are approximately 13 and
18\,km\,s$^{-1}$, respectively, we choose a minimum value of
0.1\,km\,s$^{-1}$ to provide sufficient sampling of this parameter for
unresolved spectral lines. The maximum value of 2000\,km\,s$^{-1}$ is
set by the typical maximum widths of absorption lines observed in the
literature (e.g. \citealt{Morganti:2005b}); significantly larger
values would lead to confusion with the broad baseline ripples often
present in radio spectra.

\subsubsection{Peak depth}

The maximum possible depth of a single absorption line is set by the
physical constraint that the fractional absorption cannot exceed
100\,per\,cent of the continuum flux density. By using existing
measurements of the total flux density for each radio source, we can
set an upper limit on the peak absorption depth. Reliable measurements
of the flux densities of our radio sources at either 843\,MHz and/or
1.4\,GHz are available from the SUMSS/MGPS-2 and NVSS surveys, and in
most cases their total flux densities have been compiled by
\cite{VanVelzen:2012}. However, since these continuum surveys were
carried out at different epochs to that of HIPASS, it is possible that
in some cases the continuum flux density might have varied
significantly. We therefore supplement our knowledge of the continuum
with the recently published 1.4\,GHz Continuum HIPASS (CHIPASS;
\citealt{Calabretta:2014}) compact source map. This all-sky image has
a gridded beamwidth of 14.4\,arcmin and covers the same region of sky
as the HIPASS 21\,cm line data.

To estimate the continuum flux density that was originally subtracted
from the HIPASS data, we convolve the CHIPASS image with a Gaussian
smoothing kernel, effectively reducing the spatial resolution to the
HIPASS beamwidth of 15.5\,arcmin. We then estimate the continuum flux
density within a single HIPASS beam using the weighted sum used to
generate the 21\,cm spectra (see
Section\,\ref{section:spectra_extraction}). These CHIPASS
beam-weighted flux densities for our sample of galaxies are given in
Table\,\ref{table:full_sample}. The CHIPASS beam is almost 20 times
larger than that of SUMSS, MGPS-2 and NVSS and as such we expect there
to be significant confusion with other companion radio sources.
Furthermore, for some sources the radio emission is significantly
extended beyond the HIPASS beamwidth and so we expect the CHIPASS
beam-weighted flux density to be lower than that of the total flux
density given by \cite{VanVelzen:2012}. In
Fig.\,\ref{figure:chipass_flux_comparison}, we show the CHIPASS
beam-weighted flux density versus the SUMSS/MGPS-2 and NVSS total flux
densities. This plot indicates that there is general agreement between
these quantities and the scatter is consistent with the aforementioned
issues of confusion and extension beyond the HIPASS beam. Since we
only wish to estimate the maximum possible value that the
absorption-line depth parameter can take, we use the larger of the
SUMSS/MGPS-2, NVSS and CHIPASS flux densities.  We set the lower prior
value by 1\,per\,cent of the mean noise in the spectrum, thereby
ensuring the possible detection of broad, weak absorption lines (see
e.g. \citealt{Allison:2013}) and good sampling of the depth parameter.

\subsection{Calibration error}

The flux density scale for HIPASS was calibrated using observations of
Hydra\,A and PKS\,1934-638, which have known values relative to the
absolute scale of \cite{Baars:1977}. The rms variation in the HIPASS
flux calibration was reported by \cite{Zwaan:2004} to be 2\,per\,cent
over the duration of the southern survey. If we assume that the
original flux densities obtained using the scale of \cite{Baars:1977}
have an accuracy of approximately 5\,per\,cent, then we estimate that
the HIPASS spectra should have a calibration error given by the
quadrature sum of these two errors, approximately 5.4\,per\,cent. To
propagate this error into our analysis, we introduce a parameter that
multiplies the model data at each iteration and which has a prior
probability given by a normal distribution with mean equal to unity
and $1\,\sigma$ width equal to $0.054$. In determining the
uncertainties in our model parameters, we marginalize over this
parameter.

\begin{table*} 
 \begin{threeparttable}
   \caption{A summary of derived parameters for four galaxies in which
     we have detected \mbox{H\,{\sc i}} absorption using
     HIPASS. $cz_\mathrm{sys}$ is the systemic redshift, and $S_{\rm
       1.4}$ is the 1.4\,GHz total flux density from NVSS
     \citep{Condon:1998, VanVelzen:2012}. The parameters estimated
     from model fitting are as follows: $cz_\mathrm{peak}$ is the
     21\,cm redshift at peak absorption; $\Delta{S}_{\rm peak}$ is the
     peak absorption depth; $\Delta{v_\mathrm{eff}}$ is the rest
     effective width (as defined by
     Equation\,\ref{equation:effective_width}); $\tau_\mathrm{peak}$
     is the peak optical depth, calculated using $S_{\rm 1.4}$ for the
     continuum component and assuming that the covering factor $f = 1$
     (and is therefore a lower limit to the true optical depth);
     $\int{\tau\,\mathrm{d}v}$ is the rest-frame velocity-integrated
     optical depth; $N_\mathrm{HI}$ is the \mbox{H\,{\sc i}} column
     density assuming a spin temperature of 100\,K; $R$ is the
     detection statistic as defined by
     Equation\,\ref{equation:detection_statistic}. All uncertainties
     are given for the 68.3\,per\,cent interval. The dominant source
     of uncertainty in $S_{1.4}$ for these sources is the absolute
     flux calibration error (approximately 3\,per\,cent for NVSS;
     \citealt{Condon:1998}). Given that HIPASS and NVSS are both
     ultimately calibrated to the scale of \citet{Baars:1977}, we
     assume a strong correlation with the variance in $\Delta{S}_{\rm
       peak}$ and so do not propagate the uncertainty in $S_{1.4}$
     through to our estimate of the peak optical depth and its
     dependent quantities.}\label{table:HIPASS_results}
   \setlength\tabcolsep{3pt} \renewcommand*\arraystretch{1.2}
   \begin{tabular}{lrrrrrrrrr}
     \hline
     \multicolumn{1}{c}{Name} & \multicolumn{1}{c}{$cz_\mathrm{sys}$} & \multicolumn{1}{c}{$S_{\rm 1.4}$}  & \multicolumn{1}{c}{$cz_\mathrm{peak}$} & \multicolumn{1}{c}{$\Delta{S}_{\rm peak}$} & \multicolumn{1}{c}{$\Delta{v_\mathrm{eff}}$} & \multicolumn{1}{c}{$\tau_{\rm peak}$} & \multicolumn{1}{c}{$\int{\tau\,\mathrm{d}v}$} & \multicolumn{1}{c}{$N_{\rm HI}$} & \multicolumn{1}{c}{$R$} \\
     & \multicolumn{1}{c}{(km\,s$^{-1}$)} & \multicolumn{1}{c}{(mJy)} & \multicolumn{1}{c}{(km\,s$^{-1}$)} & \multicolumn{1}{c}{(mJy)} & \multicolumn{1}{c}{(km\,s$^{-1}$)} & & \multicolumn{1}{c}{(km\,s$^{-1}$)} & \multicolumn{1}{c}{($10^{20}$\,cm$^{-2}$)} & \\
     \hline
     2MASX\,J13084201-2422581 & $4257\pm45$ & $474\pm14$ & $4224.4_{-2.7}^{+2.7}$ & $109_{-10}^{+11}$ & $71.0_{-6.7}^{+6.9}$ & $0.26_{-0.03}^{+0.03}$ & $17.8_{-1.7}^{+1.8}$ & $32.4_{-3.0}^{+3.4}$ & $104.6\pm0.1$ \\
     Centaurus\,A & $547\pm5$ & $4520\pm226$\tnote{$a$} & $548.7_{-0.5}^{+0.6}$ & $1690_{-550}^{+770}$ & $7.4_{-2.3}^{+3.6}$ & $0.47_{-0.18}^{+0.32}$ & $3.1_{-0.2}^{+0.3}$ & $5.7_{-0.4}^{+0.5}$ & $898.9\pm0.3$ \\
     NGC\,5793 & $3491\pm66$ & $1200\pm36$ & $3516.6_{-1.4}^{+1.2}$ & $880_{-47}^{+48}$ & $109.3_{-2.1}^{+2.1}$ & $1.32_{-0.14}^{+0.16}$ & $116_{-9}^{+11}$ & $211_{-17}^{+19}$ & $11\,090.0\pm0.3$ \\
     Arp\,220 & $5434\pm7$ & $326\pm10$ & $5477_{-15}^{+16}$ & $42.5_{-4.5}^{+4.7}$ & $375_{-41}^{+46}$ & $0.14_{-0.02}^{+0.02}$ & $51.2_{-5.5}^{+6.2}$ & $93_{-10}^{+11}$ & $77.0\pm0.1$ \\
     \hline
   \end{tabular}
   \begin{tablenotes}
   \item[$a$] {Centaurus A has significantly extended 1.4\,GHz
       continuum emission with respect to the HIPASS beamwidth and so
       we use the core flux density measured by \citet{Tingay:2003} to
       estimate the peak optical depth and its dependent
       quantities. We assume that the dominant source of uncertainty
       for this measurement is the absolute flux calibration error of
       5\,per\,cent given by Tingay et al.}
   \end{tablenotes}
 \end{threeparttable}
\end{table*}

\subsection{Derived quantities}

\begin{figure*}
\centering
\includegraphics[width = 0.85\textwidth]{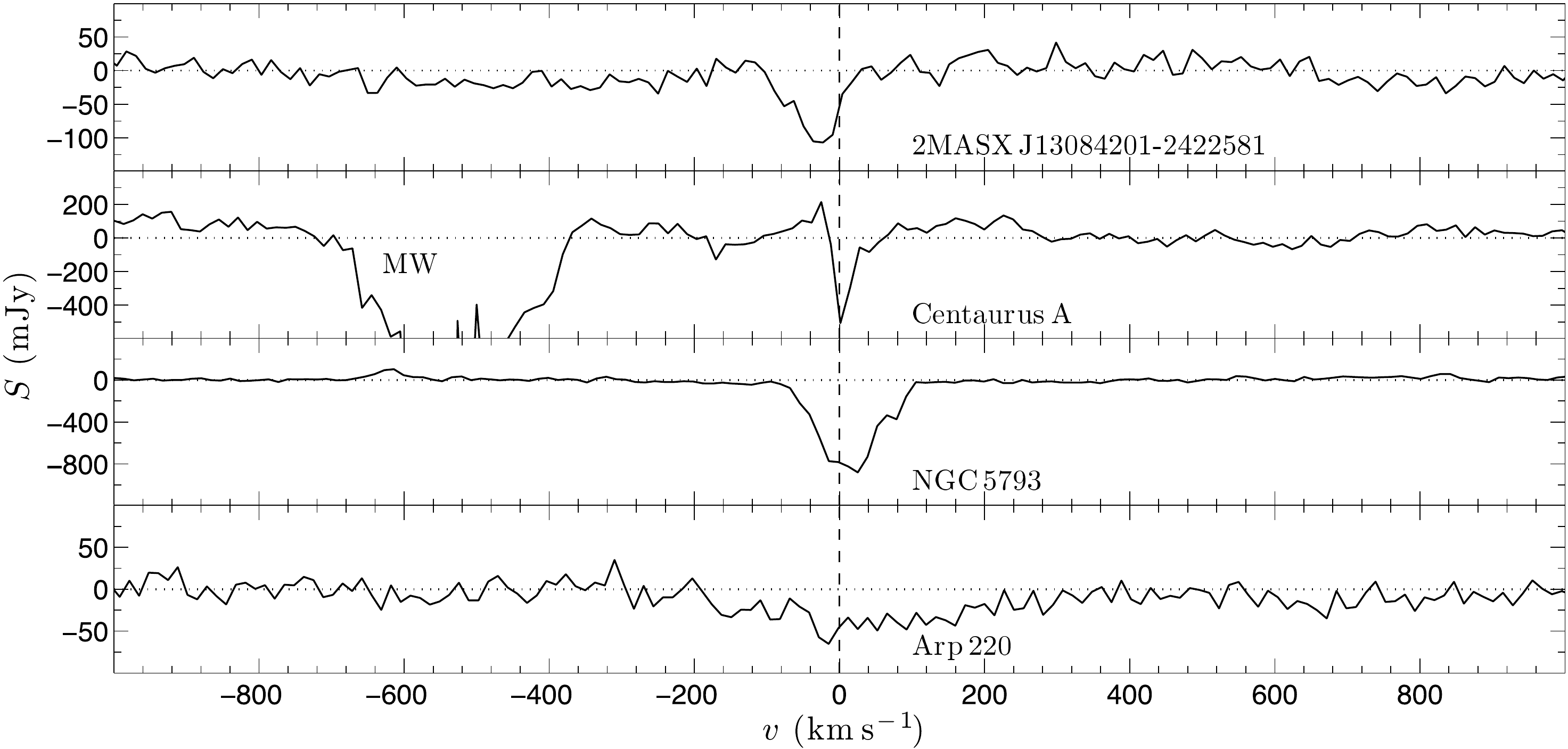}
\caption{HIPASS spectra at the position of four galaxies in which we
  have detected absorption. The radial velocity axis is given relative
  to the rest frame defined by the systemic redshift of the host
  galaxy (see Table\,\ref{table:full_sample}). The absorption in
  2MASX\,J13084201-2422581 was previously unknown. The broad feature
  ($v \sim -600$\,km\,s$^{-1}$) towards Centaurus\,A is consistent
  with being Galactic in origin.}\label{figure:hipass_detections}
\end{figure*}

\begin{figure*}
\centering
\includegraphics[width = 0.8\textwidth]{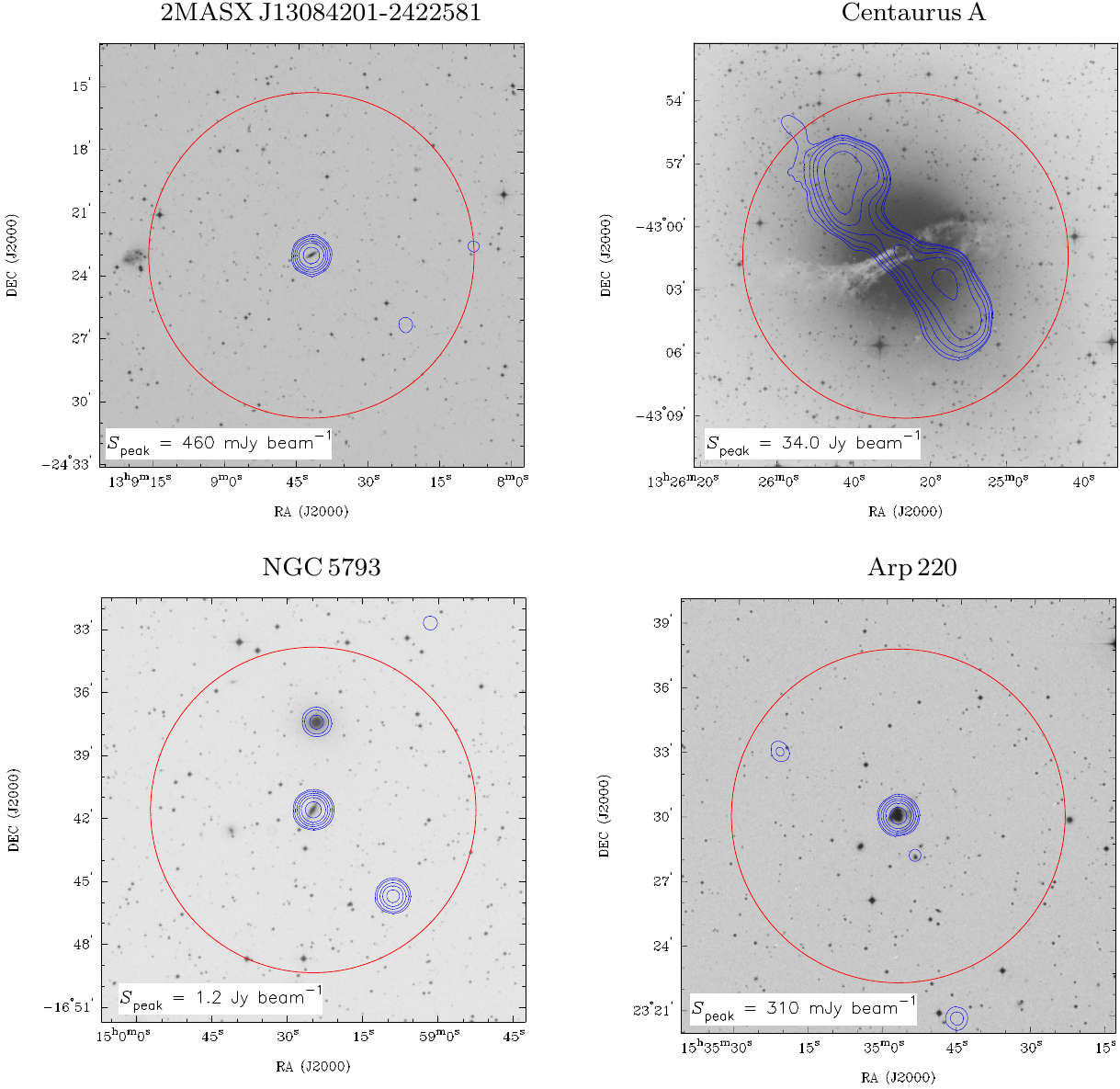}
\caption{Radio and optical images of the four galaxies in which we
  have detected absorption using HIPASS. The blue contours denote the
  1, 2, 5, 10, 20, 50\,per\,cent levels of the peak radio flux density
  within the HIPASS beamwidth, from either NVSS ($\delta > -30\degr$)
  or SUMSS/MGPS-2 ($\delta < -30\degr$). For clarity we exclude those
  radio contours that are less than five times the survey rms. The
  large red circle represents the gridded HIPASS beamwidth of
  15.5\,arcmin. The grey-scale images represent optical $B_{j}$-band
  photometry from the SuperCosmos Sky Survey, using the UK Schmidt and
  Palomar Oschin Schmidt telescopes \citep{Hambly:2001}. }
\label{figure:image_overlays_detections}
\end{figure*}

Model parametrization allows us to estimate those properties of the
absorption that we are interested in. In the regime where the
background source is significantly brighter than foreground
\mbox{H\,{\sc i}} emission, the 21\,cm optical depth across the line
profile, $\tau(v)$, can be recovered from the absorption ($\Delta{S}$)
of the continuum ($S_\mathrm{cont}$) by
\begin{equation}\label{equation:optical_depth}
  \tau(v) = -\ln\left[1 - {1\over f}\left({\Delta{S(v)}\over{S_\mathrm{cont}(v)}}\right)\right],
\end{equation}
where $f$ (the covering factor) is the fractional projected area of
continuum obscured by the absorbing gas, and $v$ is the velocity with
respect to the rest frame of the system. It should be noted that
throughout this work we assume that $f = 1$, so that estimates of
$\tau$ are a lower limit to the true optical depth. The column density
of \mbox{H\,{\sc i}} gas ($N_{\rm HI}$, in units of ${\rm cm}^{-2}$)
can be estimated from the velocity- integrated optical depth (in units
of ${\rm km\,s}^{-1}$) using the following relationship
\citep[e.g.][]{Wolfe:1975},
\begin{equation}\label{equation:column_density}
  N_\mathrm{HI} = 1.823 \times 10^{18}~T_\mathrm{spin}~\int{\tau(v)~\mathrm{d}v},
\end{equation}
where the spin temperature, $T_\mathrm{spin}$ (in units of K), is the
excitation temperature for the 21\,cm transition and hence a measure
of the relative populations of the two hyperfine states of the
hydrogen 1s ground level. $T_\mathrm{spin}$ is determined by both
radiative and collisional processes, converging to the kinetic
temperature for a collision dominated gas (e.g. \citealt{Purcell:1956,
  Field:1958, Field:1959b}).

For the purpose of comparing the widths of absorption lines, we define
the rest effective width (see also \citealt{Dickey:1982} and
\citealt{Allison:2013}) as
\begin{equation}\label{equation:effective_width}
  \Delta{v}_\mathrm{eff} \equiv \left[\int{{\Delta{S}(v)\over{S_\mathrm{cont}(v)}} ~ \mathrm{d}v}\right] \Big/ \left({\Delta{S}\over{S_\mathrm{cont}}}\right)_\mathrm{\!peak},
\end{equation}
where $v$ is the rest-frame radial velocity (referenced with respect
to the systemic redshift) and
$(\Delta{S}/S_\mathrm{cont})_\mathrm{peak}$ is the peak fractional
absorption. This quantity has advantages over both the FWHM and the
full-width at zero intensity, since it is more representative of the
width of complex multicomponent line profiles, which might have broad
and shallow wings, and is not as strongly influenced as by the S/N.

\subsection{Detections}

Using the automated method outlined above, we obtain 51 potential
detections of absorption-like features in our 204 HIPASS
spectra. Further visual inspection of all the spectra confirms that 47
are likely to be false positives, which in some cases were rejected
due to their low significance ($R \lesssim 1$) relative to the
continuum-only hypothesis. However, the majority are found to be
associated with negative features generated by spectral baseline
ripples, which are significant compared to the noise. It is clear from
these results that when such strong spectral baseline ripples are
present, the most effective and robust methods of absorption-line
detection are to either use an automated method followed by visual
inspection, as was done here, or to account for the effect of these
nuisance signals a priori using the covariance matrix. After rejecting
these false positives, we are left with four detections that we
classify as real \mbox{H\,{\sc i}} absorption lines, associated with
four nearby galaxies.

In Table\,\ref{table:HIPASS_results}, we summarize the \mbox{H\,{\sc
    i}} parameters derived from model fitting to the HIPASS
spectra. In Figs\,\ref{figure:hipass_detections} and
\ref{figure:image_overlays_detections}, we show the spectra and
images, respectively, at the positions of the four galaxies. Of these
detections, three were previously known: Centaurus\,A
\citep{Roberts:1970}, NGC\,5793 \citep{Jenkins:1983} and Arp\,220
\citep{Mirabel:1982}, while the fourth, 2MASX\,J13084201-2422581, was
not previously reported in the literature. It should be noted that
while the first three galaxies are common to both samples,
2MASX\,J13084201-2422581 is only listed in Sample 1, since its 2MASS
$K_{s}$-band magnitude ($K_{s} = 11.8$) was not bright enough to be
included in the catalogue of \cite{VanVelzen:2012}. We discuss further
the results of the model parametrization and inferred properties of
the \mbox{H\,{\sc i}} absorption in
Section\,\ref{section:individual_sources}.

\section{Followup with the ATCA}

\subsection{Observations}

We examined further the 21\,cm absorption seen in
2MASX\,J13084201-2422581 and NGC\,5793 by re-observing these galaxies
with the Australia Telescope Compact Array (ATCA) Broadband Backend
\citep[CABB;][]{Wilson:2011} in 2013 February 13--16. Our aims were
twofold: to confirm the new detection of \mbox{H\,{\sc i}} absorption
in 2MASX\,J13084201-2422581 and to verify the tentatively detected
broad absorption wings seen towards NGC\,5793 by
\citet{Koribalski:2012}.

Observations were carried out in a similar manner to those reported by
\citet{Allison:2012a,Allison:2013}. We used the 64\,MHz zoom band
capability of CABB to position 2048 spectral channels (with velocity
resolution $\sim6.7$\,km\,s$^{-1}$) at a centre frequency of
1406\,MHz, equivalent to 21\,cm redshifts in the range $ -3\,670
\lesssim cz \lesssim 10\,130$\,km\,s$^{-1}$. This band provides almost
three times the spectral resolution of HIPASS, and comfortably
includes the redshifts of the two galaxies. The six-element ATCA was
arranged in the 6A east-west configuration with baselines in the range
0.337--5.939\,km. At 1406\,MHz, this configuration provides an angular
scale sensitivity range of 7--130\,arcsec and a primary beam FWHM of
approximately 35\,arcmin. Short scans of the target fields were
interleaved with regular observations of nearby bright point sources
for gain calibration (PKS\,1308-220 and PKS\,1504-166), with a total
on-target integration time of 2\,h 30\,min for
2MASX\,J13084201-2422581 and 2\,h 15\,min for NGC\,5793. We observed
PKS\,1934-638 for calibration of the band-pass and absolute flux
scale.

\subsection{Data reduction}

The ATCA data were flagged, calibrated and imaged in the standard way
using tasks from the \textsc{Miriad}
package\footnote{http://www.atnf.csiro.au/computing/software/miriad}
\citep{Sault:1995}. Manual flagging was performed using the task
\textsc{Uvflag} for known radio frequency interference (RFI) in 80
channels at 1381\,MHz (from mode L3 of the Global Positioning System)
and 60 channels at 1431\,MHz (from the 1.5\,GHz terrestrial microwave
link band), as well as 20 channels for the 1420\,MHz Galactic 21\,cm
signal. The remaining 1888 channels were automatically flagged for
transient glitches and low-level RFI using iterative calls to the task
\textsc{Mirflag}, resulting in less than 2\,per\,cent of the data per
channel being lost. Initial calibration was performed using bright
calibrator sources, to correct the band-pass, gains and absolute flux
scale. Further correction of the gain phases was performed using
self-calibration based on a continuum model of each target
field. Continuum models were generated using the multi-frequency
deconvolution task, \textsc{Mfclean}, which recovers both the fluxes
and spectral indices of the brightest sources in the field. The NVSS
catalogue ($S_{1.4} \gtrsim 2.5$\,mJy; \citealt{Condon:1998}) was used
to identify the positions of these sources.

\begin{figure*}
\centering
\includegraphics[width = 1.0\textwidth]{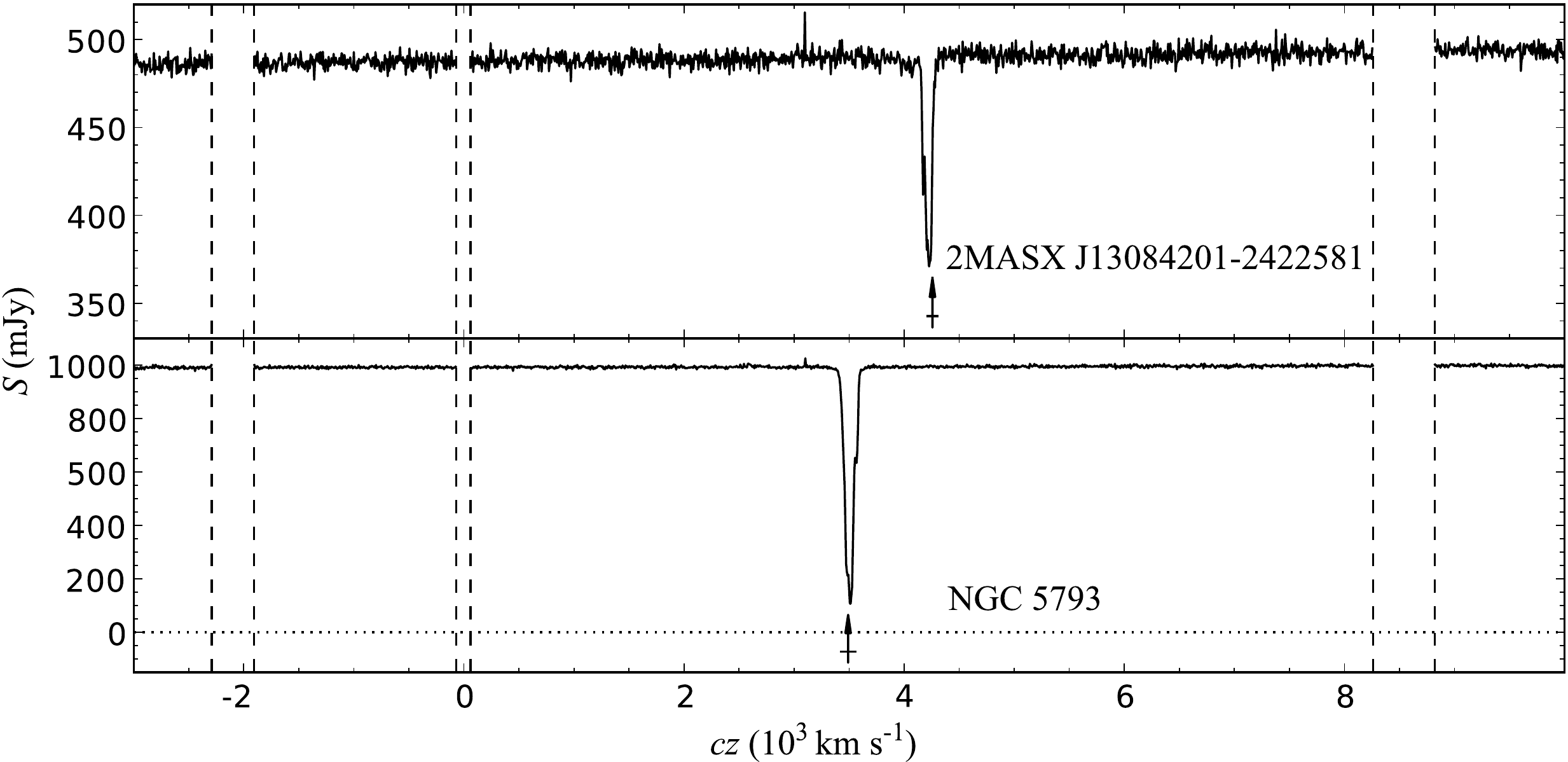}
\caption{CABB 21\,cm spectra showing both continuum emission and
  \mbox{H\,{\sc i}} absorption at the positions of radio sources
  hosted by 2MASX\,J13084201-2422581 and NGC\,5793. The arrow and
  horizontal line indicate the mean and uncertainty in the systemic
  redshift. The vertical dashed lines enclose those spectral channels
  that were flagged either due to persistent strong RFI or Galactic
  21\,cm signal. The emission spike that is apparent in both spectra,
  within a single channel at the band centre (1406\,MHz, $cz \approx
  3100\,\mathrm{km}\,\mathrm{s}^{-1}$), arises due to self-generated
  interference within the telescope (see
  \citealt{Wilson:2011}).}\label{figure:cabb_spectra}
\end{figure*}

Initially, we imaged the target fields by uniformly weighting the
calibrated visibilities, thereby favouring the undersampled longer
baselines and so optimizing the spatial resolution. However, from
visual inspection of these uniformly weighted images, we found that
the target sources are only resolved on scales smaller than the
synthesized beam FWHM of $\sim$\,10\,arcsec. Based on this
information, we instead used the natural weighting scheme to generate
our final continuum and spectral images, which optimizes the S/N for
detection. Before constructing our final data cubes, we subtracted a
continuum model of other nearby sources in the field, thereby removing
significant spectral baseline artefacts generated from incomplete
Fourier sampling. A spectrum was then extracted from each data cube at
the position of the target source, using the method described in
Section\,\ref{section:spectra_extraction} for an elliptical
synthesized beam. In Table\,\ref{table:atca_observations}, we
summarize some properties of our ATCA observations and in
Figs\,\ref{figure:cabb_spectra} and \ref{figure:image_overlays_cabb}
we show the final CABB spectra and images, respectively, for both
targets. Note that our analysis method does not require any smoothing
of the spectral data.

\begin{table} 
  \caption{A summary of our ATCA observations, 
    where $t_{\rm int}$ is the total integration time on each source; $\theta_{\rm maj}$, 
    $\theta_{\rm min}$ and $\phi$ are the major axis, minor axis and 
    position angle, respectively, of an elliptical fit to the synthesized beam; and 
    $\sigma_\mathrm{chan}$ is the per-channel noise estimate in 
    the CABB spectra.}\label{table:atca_observations}
  \setlength\tabcolsep{3pt} \renewcommand*\arraystretch{1.0}
 \begin{tabular}{@{}lrrrrr}
   \hline
   \multicolumn{1}{c}{Name} & \multicolumn{1}{c}{$t_{\rm int}$} & \multicolumn{1}{c}{$\theta_{\rm maj}$} & \multicolumn{1}{c}{$\theta_{\rm min}$} & \multicolumn{1}{c}{$\phi$} & \multicolumn{1}{c}{$\sigma_\mathrm{chan}$}  \\
   & \multicolumn{1}{c}{(hh:mm)} & \multicolumn{2}{c}{(arcsec)} & \multicolumn{1}{c}{(deg)} & \multicolumn{1}{c}{(mJy)} \\
   \hline
   2MASX\,J13084201-2422581 & 02:30 & 19.5 & 9.2 & -5.2 & 3.2 \\
   NGC\,5793 & 02:15 & 28.6 & 8.5 & -7.9 & 3.9 \\
   \hline
   \end{tabular}
\end{table}

\begin{table*} 
\begin{threeparttable}
  \caption{A summary of derived \mbox{H\,{\sc i}} absorption
    parameters for 2MASX\,J13084201-2422581 and NGC\,5793, estimated
    from our CABB 21\,cm spectra. $S_\mathrm{cont}$ is the continuum
    flux density at the position of peak absorption and $\chi_{\rm
      ml}^{2}/{\rm d.o.f.}$ is the reduced chi-squared statistic for
    the maximum likelihood model parameters; the other parameters are
    as defined in Table\,\ref{table:HIPASS_results}. All uncertainties
    are given for the 68.3\,per\,cent
    interval.}\label{table:CABB_results}  
  \setlength\tabcolsep{2pt} \renewcommand*\arraystretch{1.2} 
 \begin{tabular}{@{}lrrrrrrrrr}
   \hline
   \multicolumn{1}{c}{Name} & \multicolumn{1}{c}{$S_\mathrm{cont}$}  & \multicolumn{1}{c}{$cz_\mathrm{peak}$} & \multicolumn{1}{c}{$\Delta{S}_{\rm peak}$} & \multicolumn{1}{c}{$\Delta{v_\mathrm{eff}}$} & \multicolumn{1}{c}{$\tau_{\rm peak}$} & \multicolumn{1}{c}{$\int{\tau\,\mathrm{d}v}$}  & \multicolumn{1}{c}{$N_\mathrm{HI}$} &  \multicolumn{1}{c}{$R$} & $\chi_{\rm ml}^{2}/{\rm d.o.f.}$\\
   & \multicolumn{1}{c}{(mJy)} & \multicolumn{1}{c}{(km\,s$^{-1}$)} & \multicolumn{1}{c}{(mJy)} & \multicolumn{1}{c}{(km\,s$^{-1}$)} & & \multicolumn{1}{c}{(km\,s$^{-1}$)} &  \multicolumn{1}{c}($10^{20}$\,cm$^{-2}$) &  & \\
   \hline
   2MASX\,J13084201-2422581 & $499_{-32}^{+35}$ & $4238.0_{-3.5}^{+2.0}$ & $122.5_{-8.0}^{+8.6}$ & $86.8_{-2.2}^{+2.2}$ & $0.28_{-0.01}^{+0.01}$ & $23.3_{-0.4}^{+0.5}$ & $42.5_{-0.8}^{+0.8}$ & $6061.8\pm0.3$ & $1.22\pm0.03$ \\
   NGC\,5793 & $1022_{-73}^{+77}$ & $3514.9_{-0.4}^{+0.4}$ & $922_{-67}^{+69}$ & $97.7_{-0.3}^{+0.3}$ & $2.32_{-0.03}^{+0.03}$ & $157.6_{-0.5}^{+0.5}$ & $287.2_{-1.0}^{+1.0}$ & $270\,809.6\pm0.3$ & $1.08\pm0.03$ \\
   \hline
  \end{tabular}
\end{threeparttable}
\end{table*}

\subsection{CABB data analysis and modelling}\label{section:cabb_analysis}

We determine best-fitting models of the \mbox{H\,{\sc i}} absorption
in each CABB spectrum using the Bayesian method described in
Section\,\ref{section:line_finding}. The continuum component is
parametrized using a first-order polynomial (linear) model and the
absorption line by the combination of multiple Gaussian
components. The best-fitting number of Gaussian components is then
optimized by maximizing the statistic $R$
(Equation\,\ref{equation:detection_statistic}). We assume that the
CABB data have a rectangular spectral point spread function, so that
the channels are independent of each other \citep{Wilson:2011} and
hence the covariance matrix $\textbfss{C}$ in
Equation\,\ref{equation:likelihood} reduces to the on-diagonal set of
channel variances, estimated using the MADFM statistic over the
spectrum.

\begin{figure*}
\vspace{6pt}
\centering
\includegraphics[width = 1.0\textwidth]{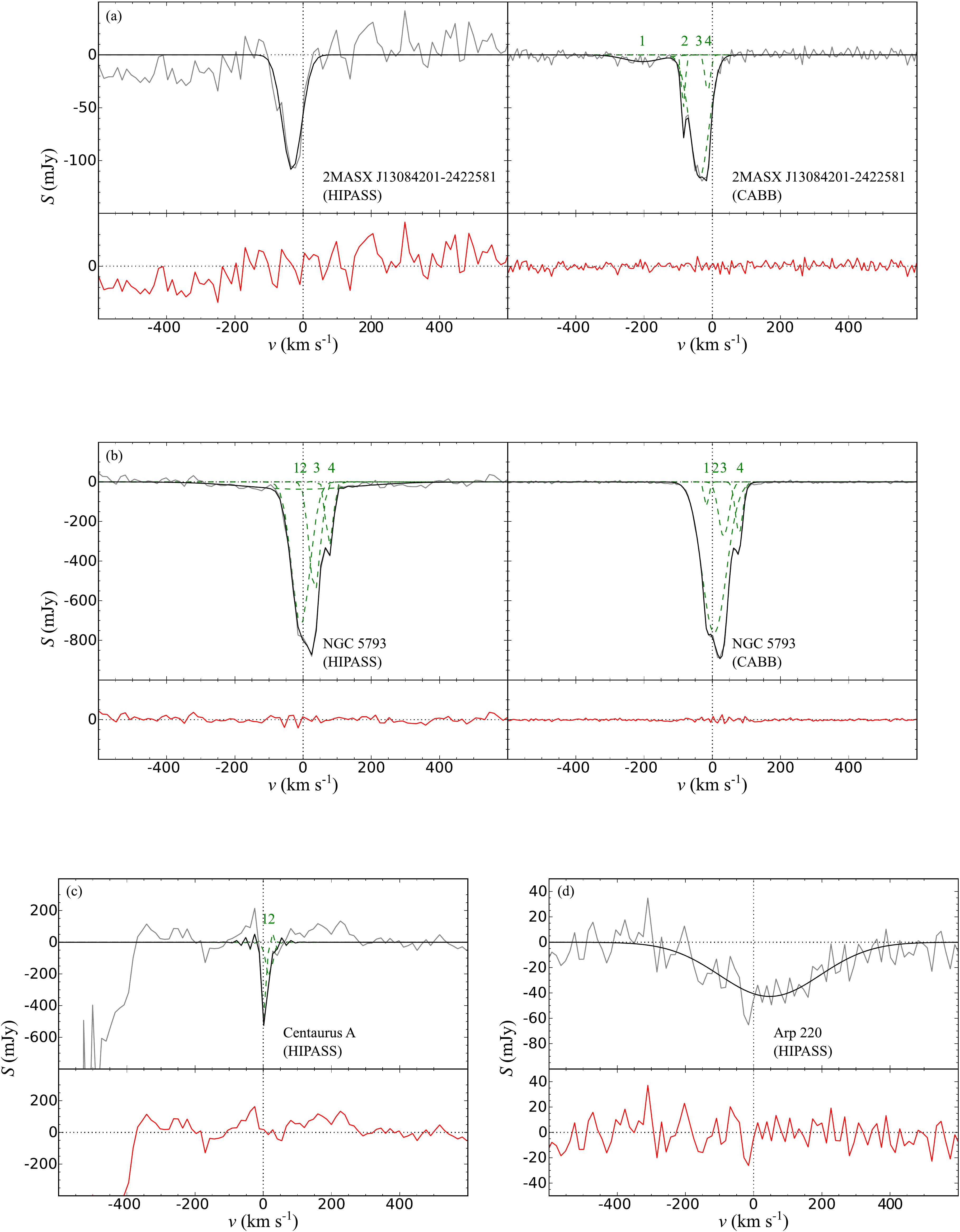}
\caption{The best-fitting models of HIPASS and CABB spectra in which
  we have detected associated \mbox{H\,{\sc i}} absorption. The radial
  velocity axis is given relative to the rest frame defined by the
  systemic redshift (vertical dotted line). For the CABB spectra, the
  grey line represents the data after we have subtracted our
  best-fitting continuum component, whereas for the HIPASS spectra the
  grey line represents the original baseline-subtracted data. The
  solid black line represents the best-fitting spectral line model and
  the dashed green lines the individual Gaussian components (top
  panel). The solid red line represents the best-fitting residual
  (bottom panel).}\label{figure:spectral_fits}
\end{figure*}

While it is reasonable to assume that the CABB spectra are free of
strong spectral baseline artefacts, we again use a normal probability
distribution for the position of the absorption line, centred on the
systemic redshift and within the 1$\sigma$ width of
$\pm50$\,km\,s$^{-1}$. We do this for two reasons: to encode our prior
belief that the absorption should arise near the known systemic
redshift, and to avoid unnecessarily fitting to any broad and shallow
spectral baseline ripples that might exist at either edge of the
spectrum. The FWHM of each Gaussian component is given a uniform prior
of 0.1--2000\,km\,s$^{-1}$, and the lower and upper limits of the
depth are set by 1\,per\,cent of the per-channel noise and the mean
continuum flux density, respectively.  We assume a systematic error of
$\pm$10\,per\,cent for the calibration procedure, which is
approximated by multiplying the model data by an additional parameter
with a normal prior of $1.0\pm0.1$. While this nuisance parameter
increases the uncertainty in our estimates of the absolute flux scale
of the continuum and spectral line components, it does not
significantly alter the relative fractional absorption. A summary of
the estimated \mbox{H\,{\sc i}} parameters from model fitting to the
CABB spectra is given in Table\,\ref{table:CABB_results}. In
Table\,\ref{table:parameters} and Fig.\,\ref{figure:spectral_fits} we
summarize the best-fitting parameters, for multiple Gaussian
components, for both the HIPASS and CABB data.

\section{Discussion}

\begin{figure*}
\centering
\includegraphics[width = 1.0\textwidth]{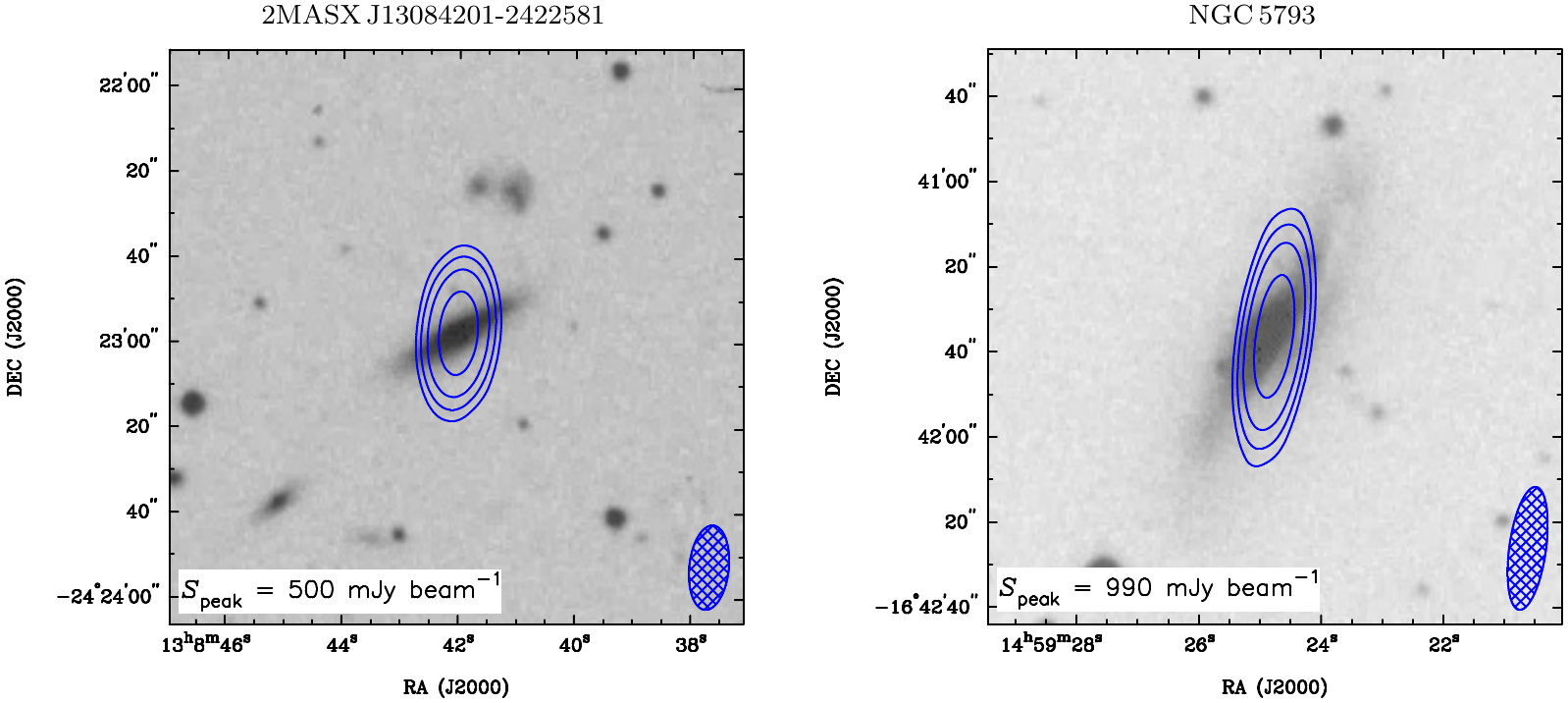}
\caption{1.4\,GHz ATCA images of 2MASX\,J13084201-2422581 and
  NGC\,5793, where the blue contours denote the 5, 10, 20,
  50\,per\,cent levels of the peak continuum flux density. The FWHM of
  the synthesized beam is shown in the bottom-right corner. The
  grey-scale images represent optical $B_{j}$-band photometry from the
  SuperCosmos Sky Survey, using the UK Schmidt and Palomar Oschin
  Schmidt telescopes \citep{Hambly:2001}.}
\label{figure:image_overlays_cabb}
\end{figure*}

\subsection{Individual sources}\label{section:individual_sources}

\subsubsection{2MASX\,J13084201-2422581}

We detect a previously unknown 21\,cm absorption line against the
compact flat-spectrum radio source at the centre of the Seyfert~2
galaxy 2MASX\,J13084201-2422581. We show in
Fig.\,\ref{figure:spectral_fits} the best fitting models to both the
HIPASS and CABB spectra. By comparing the marginal likelihoods for
increasingly complex models, we find that the HIPASS data warrant only
a single-component Gaussian model, while a four-component model is
favoured by the CABB data. The similarity in our estimates of the peak
depth and rest effective width for each spectrum implies that all of
the \mbox{H\,{\sc i}} absorption detected within the 15.5\,arcmin
HIPASS beam arises from a region of angular size smaller than the ATCA
synthesized beam, which at $cz = 4257$\,km\,s$^{-1}$ equates to a
projected physical size smaller than $6 \times 3$\,kpc. This result is
consistent with the compact morphology of the radio source at
1.4\,GHz, evident from the NVSS and ATCA images, and the absence of
other nearby strong radio sources within the HIPASS beam (see
Fig.\,\ref{figure:image_overlays_detections}).

The redshift of the peak absorption is consistent with the systemic
redshift of the host galaxy, implying that the bulk of the cold
$\mbox{H\,{\sc i}}$ gas is not rapidly infalling or outflowing with
respect to ionized gas in the nucleus. The stellar component of the
host galaxy exhibits an edge-on irregular spiral morphology at
near-infrared and optical wavelengths \citep{Jarrett:2000,
  Hambly:2001}, which, with the Seyfert~2 classification of the AGN,
suggests that the bulk of the absorbing gas may arise within an
obscuring disc of \mbox{H\,{\sc i}} gas. While we cannot spatially
resolve the background radio-jet structure with our ATCA observations,
and hence strongly constrain the spatial distribution and kinematics
of the \mbox{H\,{\sc i}} gas, we note that the shape and width of the
absorption-line profile are very similar to those observed in other
Seyfert galaxies (e.g. \citealt{Dickey:1982, Dickey:1986,
  Gallimore:1999}). Work by \citet{Gallimore:1999} showed that these
systems are well modelled by sub-kpc discs of \mbox{H\,{\sc i}} gas
that are typically aligned with the outer stellar disc. Deviations in
the regular shape of the main profile (components 2, 3 and 4 in
Fig.\,\ref{figure:spectral_fits}) are likely generated by a
combination of unresolved spatial variations in the optical depth of
the gas, the complex geometries of the absorber-radio source system
and radial streaming of the gas with respect to the source. The
separate broad and shallow blueshifted component at
$v\sim200\,\mathrm{km}\,\mathrm{s}^{-1}$ indicates that gas might be
caught in a jet-driven outflow on sub-kpc scales, but this
interpretation remains tentative until the absorption can be spatially
resolved.

Our best estimates of the peak and integrated 21\,cm optical depths
from the CABB data are $0.28\pm0.01$ and $23.3\pm0.5$\,km\,s$^{-1}$,
respectively (assuming that $f=1$). However, without further knowledge
of the relative size and geometry of the absorbing gas with respect to
the continuum source, as well as the spin temperature of gas, it is
very difficult to obtain an accurate measurement of the column density
from Equation\,\ref{equation:column_density}. \cite{Gallimore:1999}
showed that for a dense AGN-irradiated gas cloud in the narrow-line
region of a Seyfert galaxy, the 21\,cm spin temperature is likely to
be collisionally dominated with typical values of $T_\mathrm{spin} =
100$\,K. However, if the sightline to the continuum source intercepts
the warmer atomic medium, then the spin temperature may be much
higher. For example, 21\,cm observations of intervening damped Lyman
$\alpha$ absorbers ($N_\mathrm{HI} > 2 \times
10^{20}\,\mathrm{cm}^{-2}$) show that $T_\mathrm{spin}/f$ can be
greater than $1000\,\mathrm{K}$ (see \citealt{Curran:2012} and
references therein). Furthermore, if the \mbox{H\,{\sc i}} gas in this
Seyfert~2 galaxy is distributed as a disc on scales less than 100\,pc,
then it would be unlikely that all of the source structure would be
uniformly obscured by the absorbing gas, and so in this case we would
expect the covering factor to be less than unity. Therefore, given the
possible values of $T_\mathrm{spin}$ and $f$, we can only estimate a
lower limit to the \mbox{H\,{\sc i}} column density of $N_\mathrm{HI}
= 42.5\pm0.8 \times
10^{20}~(T_\mathrm{spin}/100\,\mathrm{K})~\mathrm{cm}^{-2}$. The
stellar disc evident in the 2MASS $K_{s}$-band photometry for this
galaxy has a major--minor axis ratio of 0.380 \citep{Jarrett:2000},
which we convert into an inclination angle of $i = 74\degr$
\citep{Tully:1977, Aaronson:1980}. Assuming that the \mbox{H\,{\sc i}}
gas is coplanar with the stellar component, our estimate of the column
density is consistent with the inclination angle relationship measured
in other Seyferts and active galaxies by \cite{Dickey:1982,Dickey:1986} and
\cite{Gallimore:1999}.

\begin{table}
\begin{threeparttable}
  \caption{Best-fitting parameters from fitting multiple Gaussian
    components to the HIPASS and CABB 21\,cm spectra. $n$ is the
    component number, corresponding to the label given in
    Fig.\,\ref{figure:spectral_fits}; $cz$ is the component redshift;
    $\Delta{v_{\rm FWHM}}$ is the velocity FWHM and $\Delta{S}$ is the
    depth.}\label{table:parameters} \setlength\tabcolsep{2pt}
  \setlength\tabcolsep{1pt}
  \begin{tabular}{@{}llrrrr@{}}
    \hline
    \multicolumn{1}{l}{Name} & \multicolumn{1}{l}{Data} & \multicolumn{1}{c}{$n$} & \multicolumn{1}{c}{$cz$} & \multicolumn{1}{c}{$\Delta{v_{\rm FWHM}}$} & \multicolumn{1}{c}{$\Delta{S}$} \\
    & & & \multicolumn{1}{c}{(km\,s$^{-1}$)} & \multicolumn{1}{c}{(km\,s$^{-1}$)} & \multicolumn{1}{c}{(mJy)} \\
    \hline
    2MASX\,J13084201-2422581 & HIPASS & 1 & 4224 & 66 & 102 \\
    & CABB & 1 & 4050 & 127 & 6 \\
    & & 2 & 4170 & 12 & 53 \\
    & & 3 & 4219 & 65 & 114 \\
    & & 4 & 4243 & 18 & 36 \\ 
    Centaurus\,A & HIPASS & 1 & 548 & 3 & 2686 \\
    & & 2 & 564 & 21 & 241 \\
    NGC\,5793 & HIPASS & 1 & 3484 & 63 & 702 \\
    & & 2 & 3486 & 380 & 36 \\
    & & 3 & 3526 & 39 & 544 \\
    & & 4 & 3569 & 25 & 325 \\
    & CABB & 1 & 3474 & 15 & 143 \\
    & & 2 & 3496 & 83 & 848 \\
    & & 3 & 3525 & 35 & 310 \\
    & & 4 & 3570 & 24 & 298 \\
    Arp\,220 & HIPASS & 1 & 5482 & 348 & 42 \\
    \hline
  \end{tabular}
\end{threeparttable}
\end{table}

\begin{figure}
\centering
\includegraphics[width = 1.0\columnwidth]{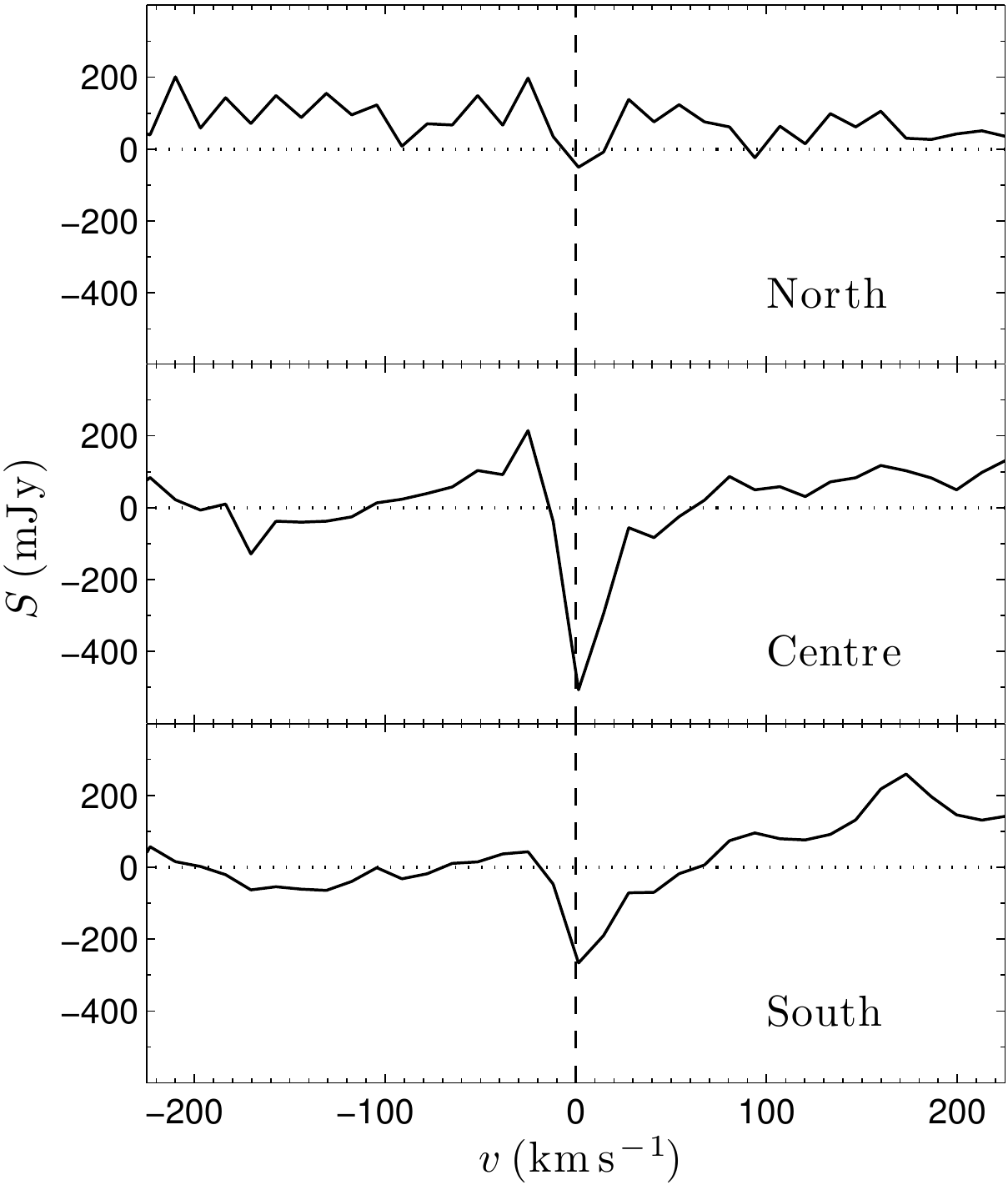}
\caption{HIPASS spectra towards Centaurus\,A, extracted at
  15.5\,arcmin separations along the observed jet-axis, at RA (J2000)
  = $13^{\rm h}26^{\rm m}27\fs72$, Dec. (J2000) =
  $-42\degr50\arcmin09\farcs8$ (north), RA (J2000) = $13^{\rm
    h}25^{\rm m}27\fs76$, Dec. (J2000) = $-43\degr01\arcmin07\farcs4$
  (centre) and RA (J2000) = $13^{\rm h}24^{\rm m}27\fs80$,
  Dec. (J2000) = $-43\degr12\arcmin05\farcs0$ (south). The radial
  velocity axis is given relative to the rest frame defined by the
  systemic redshift (vertical dashed
  line).}\label{figure:cenA_compare}
\end{figure}

\subsubsection{Centaurus\,A}

Centaurus\,A (NGC\,5128) is by far the closest early-type radio galaxy
to the Milky Way, which at a distance of only 3.8\,Mpc
\citep{Harris:2010} has been imaged in detail at multiple wavelengths
(see \citealt{Israel:1998} and references therein). In the HIPASS
spectrum, extracted from the core of the radio source, we re-detect
the \mbox{H\,{\sc i}} absorption first discovered by
\cite{Roberts:1970} and studied extensively since
\citep[e.g.][]{Whiteoak:1971, VanderHulst:1983, Sarma:2002,
  Morganti:2008, Struve:2010c}. Due to the proximity and radio power
($P_{1.4} \approx 5 \times 10^{24}$\,W\,Hz$^{-1}$) of this source, the
HIPASS spectrum is strongly contaminated by the spectral baseline
ripple. Despite this, we recover a two-component Gaussian model of the
line (Fig.\,\ref{figure:spectral_fits}c), with an effective width of
$7.4_{-2.3}^{+3.6}$\,km\,s$^{-1}$ and a peak depth of
$1690_{-550}^{+770}$\,mJy. The poor constraints on these parameter
estimates, compared with those for the other detections, are the
result of low spectral sampling across the line.

The structure of the line profile, with a deep narrow component at the
systemic redshift and an broadened component towards higher redshifts,
is consistent with the structure seen at similar spectral and spatial
resolution by \cite{Roberts:1970} and \cite{Whiteoak:1971}. The deeper
narrow component is thought to arise in absorption from \mbox{H\,{\sc
    i}} gas in a rotating disc that is coplanar with the prominent
warped dust lane, while some of the redshifted absorption is
consistent with infalling clouds towards the nucleus
\citep{VanderHulst:1983, Sarma:2002}. Higher spatial resolution and
more sensitive observations by \cite{Morganti:2008} and
\cite{Struve:2010c} revealed blueshifted absorption towards the
nucleus, potentially indicating the presence of a circumnuclear disc
of \mbox{H\,{\sc i}} on sub-100\,pc scales.

The 1.4\,GHz emission from Centaurus\,A is moderately extended with
respect to the HIPASS beam, and so in Fig.\,\ref{figure:cenA_compare}
we show spectra extracted at three positions along the observed jet
axis, separated by intervals of 15.5\,arcmin. While evidently
contaminated by residual spectral baseline signal, and subject to
adjacent signal entering from the beam sidelobes, we tentatively see
more absorption towards the southern end of the jet axis. This is
consistent with the orientation of the \mbox{H\,{\sc i}} disc against
the southern radio lobe, seen at higher spatial resolution
\citep[e.g.][]{Struve:2010c}.

\subsubsection{NGC\,5793}

The very deep 21\,cm absorption seen towards the compact and
radio-luminous nucleus in this edge-on disc Seyfert~2 galaxy was first
detected by \cite{Jenkins:1983} using the 64\,m Parkes Radio
Telescope, and has since been studied at higher spatial resolution by
\cite{Gardner:1986} using the Very Large Array (VLA), and by
\cite{Gardner:1992} and \cite{Pihlstrom:2000} using very long baseline
interferometry (VLBI). The absorption-line profile seen in the HIPASS
spectrum \citep{Koribalski:2004} is consistent with that observed by
\cite{Jenkins:1983}. The weaker emission-line feature seen in both
spectra at $cz \approx 2860$\,km\,s$^{-1}$ is attributed by
\cite{Koribalski:2012} to \mbox{H\,{\sc i}} gas in the neighbouring
dwarf irregular galaxy 6dF\,J1459410-164235 (to the east), and not the
E0 galaxy NGC\,5796 (to the north), which is thought to be relatively
\mbox{H\,{\sc i}} poor.

\begin{figure*}
\vspace{6pt}
\centering
\includegraphics[width = 1.0\textwidth]{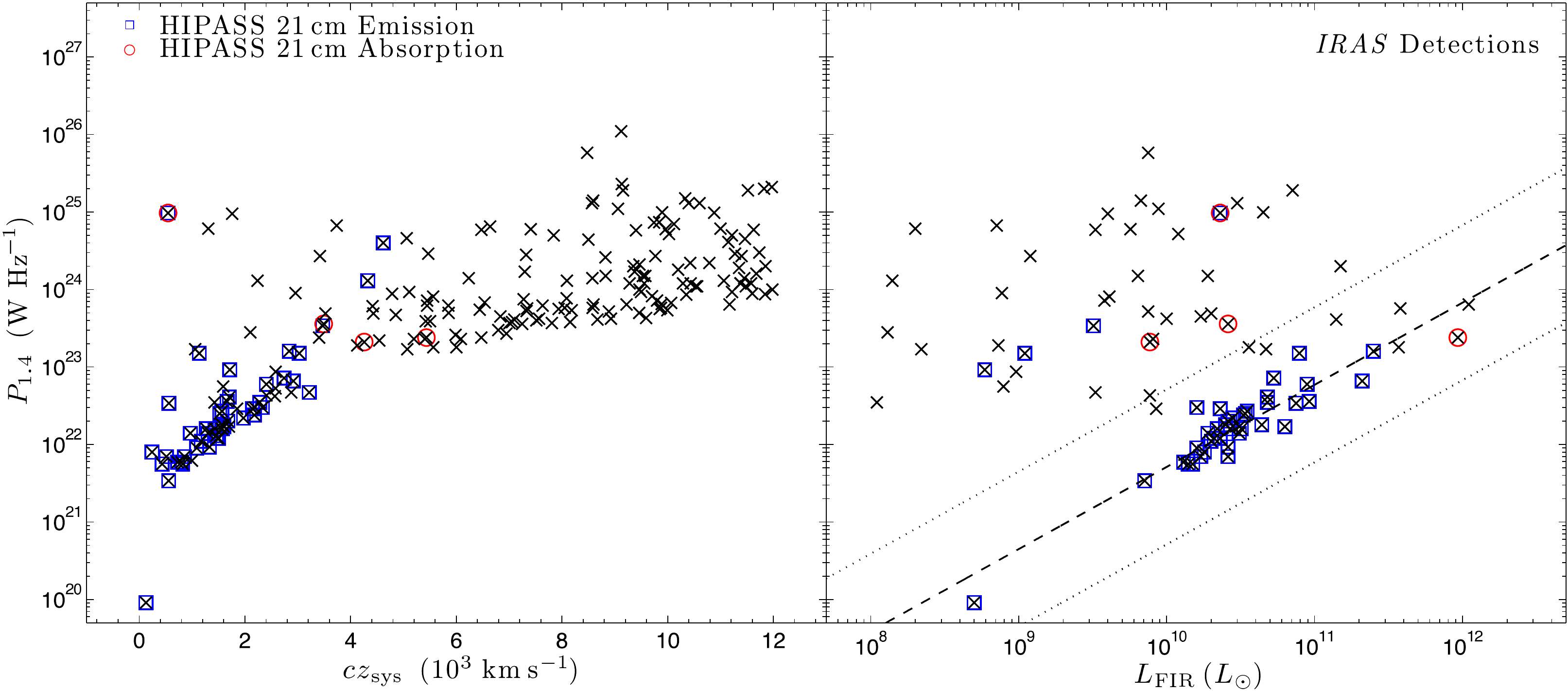}
\caption{Left: the 1.4\,GHz radio power versus the systemic redshift
  for all 204 galaxies in our sample. Right: the 1.4\,GHz radio power
  versus the far-infrared luminosity for those 86 galaxies with
  far-infrared 60 or 100\,$\mathrm{\mu}$m flux densities from the
  \emph{IRAS} survey. The blue squares represent those galaxies with
  \mbox{H\,{\sc i}} emission catalogued in HIPASS and the red circles
  denote those in which we have detected absorption. The dashed line
  shows the radio power versus far-infrared luminosity relationship
  measured by \citet{Mauch:2007} for star-forming galaxies. The dotted
  lines indicate a one-decade deviation from this
  relationship.}\label{figure:luminosity}
\end{figure*}

The spatially unresolved absorption lines in both our HIPASS and CABB
spectra clearly exhibit some velocity structure, which we successfully
model using a four-component Gaussian model (see
Fig.\,\ref{figure:spectral_fits}b). VLBI observations by
\cite{Pihlstrom:2000} demonstrated that this structure results from
the superposition of individual \mbox{H\,{\sc i}} components seen
against two continuum sources that are only resolved on angular scales
smaller than 10\,mas. They suggest that the broadest feature likely
arises in a nearly edge-on disc of \mbox{H\,{\sc i}} gas ($i \approx
73\,\degr$), and occurs on scales of 50--100\,pc from the AGN,
consistent with that seen in other Seyfert~2s \citep{Gallimore:1999},
while the other features are signatures of individual \mbox{H\,{\sc
    i}} clouds that are either interior or exterior to this disc. By
fitting a four component Gaussian model to the CABB spectrum, we
estimate that the peak and integrated optical depth are $2.32\pm0.03$
and $157.6\pm0.5$\,km\,s$^{-1}$, respectively, giving a total
\mbox{H\,{\sc i}} column density of $N_\mathrm{HI} = 287.2\pm1.0
\times 10^{20} (T_\mathrm{spin}/100\,\mathrm{K})$\,cm$^{-2}$. This is
consistent with the total column density measured by
\cite{Pihlstrom:2000}, averaged across the resolved continuum
components, of $N_\mathrm{HI} \approx 3.5 \times 10^{22}
(T_\mathrm{spin}/100\,\mathrm{K})$\,cm$^{-2}$.

\cite{Koribalski:2012} tentatively identified a previously undetected
broad absorption feature in the HIPASS spectrum, with a width of
$680$\,km\,s$^{-1}$ and centred on the systemic redshift. Using
Bayesian model comparison, we confirm that this feature is
statistically significant above the noise (component 1 in
Fig.\,\ref{figure:spectral_fits}b and Table\,\ref{table:parameters});
however, it is not clear if this feature is distinguishable from other
residual baseline features seen in the spectrum. Furthermore, we do
not re-detect this broad component in the CABB spectrum (with an
estimated per-channel noise of $\sigma_{\rm chan} = 3.9$\,mJy), even
though there is strong consistency between the other absorption
components seen in both spectra. It is plausible that this feature
could have arisen towards a confused source within the HIPASS beam,
from \mbox{H\,{\sc i}} gas that is at a similar redshift to
NGC\,5793. There are two other sources within the HIPASS beam that
have sufficient flux densities ($S_{1.4} \gtrsim 35$\,mJy) in the NVSS
catalogue to produce such an absorption: NGC\,5796 ($S_{1.4} =
109$\,mJy, $cz_{\rm sys} = 2971$\,km\,s$^{-1}$; \citealt{Wegner:2003})
and MRC\,1456-165 ($S_{1.4} = 379$\,mJy).  However, spectra extracted
from the CABB data at the centroid positions of both sources show no
evidence of the broad absorption seen in the HIPASS spectrum. We
therefore conclude that this feature is likely an artefact and the
result of residual spectral baseline ripple in the HIPASS spectrum.

\subsubsection{Arp\,220}

The broad absorption-line associated with this prototypical
ultraluminous infrared galaxy (\citealt{Sanders:2003}) was originally
detected by \cite{Mirabel:1982}, using the 300\,m Arecibo Telescope,
and has since been re-observed and studied multiple times
(e.g. \citealt{Dickey:1986, Baan:1987, Garwood:1987, Baan:1995,
  Hibbard:2000, Mundell:2001}). We find that the line detected in the
HIPASS spectrum requires only a single-component Gaussian model, with
an effective width ($\Delta{v_{\rm eff}} =
375_{-41}^{+46}$\,km\,s$^{-1}$) and peak depth ($\Delta{S_{\rm peak}}
= 42.5_{-4.5}^{+4.7}$\,mJy) that are consistent with previous
single-dish observations (e.g \citealt{Mirabel:1982,
  Garwood:1987}). However, the lower S/N and spectral resolution of
the HIPASS spectrum means that we do not find as much structure in the
line as seen in these other single-dish observations.

The absorption arises from gas towards a compact radio-loud nucleus
that consists of two distinct components (e.g. \citealt{Baan:1987,
  Norris:1988, Baan:1995}), which are thought to be the nuclei of two
gas-rich progenitor galaxies in an advanced stage of merging. At
1.4\,GHz, they are only resolved on angular scales smaller than
$\sim200$\,mas \citep[][]{Mundell:2001} and are therefore not resolved
by HIPASS. Mundell et al. carried out a high spatial resolution study
of the 21\,cm line on sub-arcsec scales, using the Multi-Element
Radio-Linked Interferometer Network array, and showed that the bulk of
the absorption is likely associated with two counterrotating discs of
\mbox{H\,{\sc i}} gas centred on each of the nuclei, consistent with
observations of emission from the CO gas content
\citep{Sakamoto:1999}. The broad width of the absorption line seen in
the HIPASS spectrum is consistent with the superposition of these
rotating components and the bridge of \mbox{H\,{\sc i}} gas connecting
the two nuclei.

\begin{figure}
\vspace{6pt}
\centering
\includegraphics[width = 1.0\columnwidth]{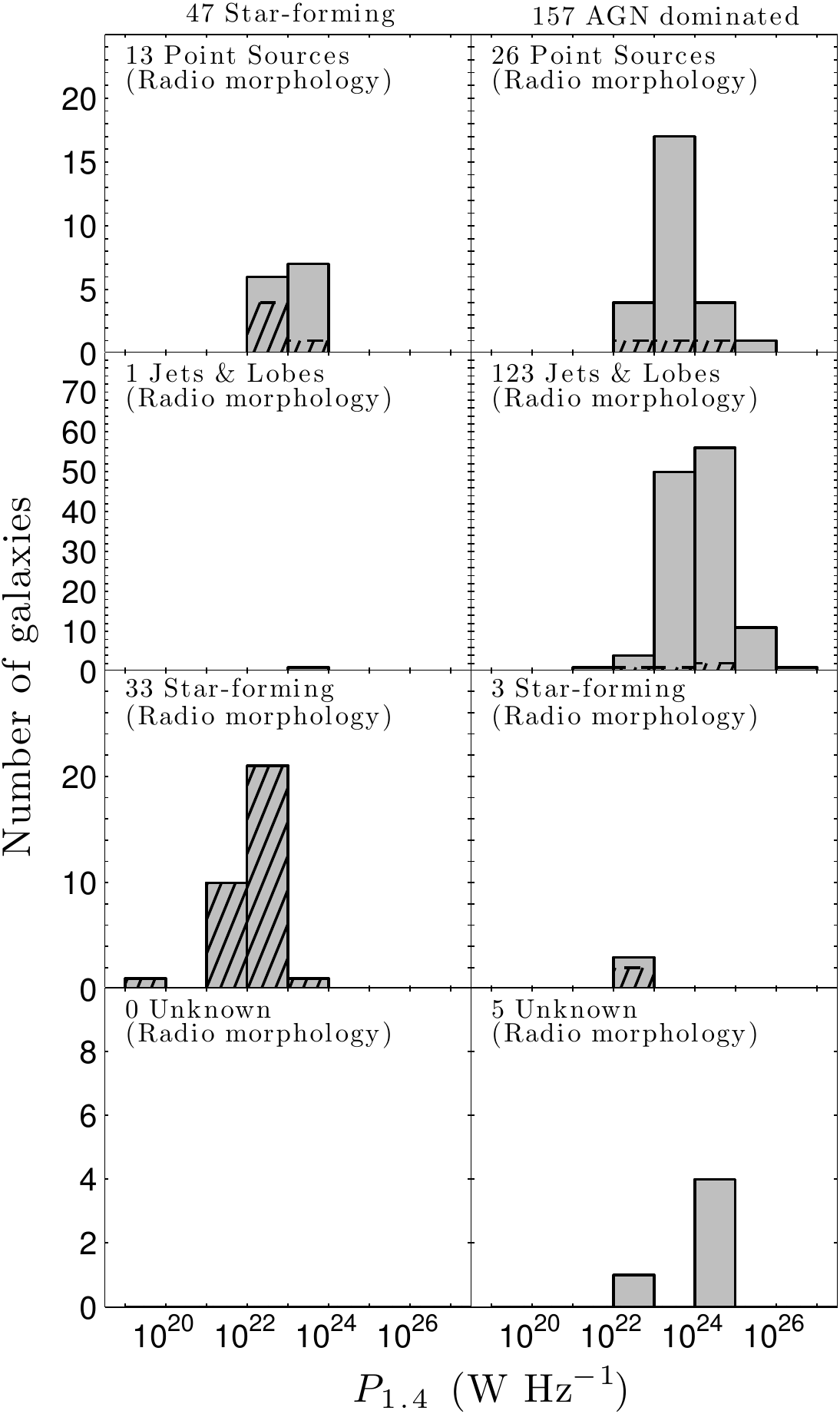}
\caption{The distribution of 1.4\,GHz radio powers for our sample,
  where we have classified galaxies as either star forming or AGN
  dominated (based on the radio and far-infrared luminosity
  relationship), and which have radio morphologies as classified by
  \citet{VanVelzen:2012}. The hatched bars represent those galaxies
  with catalogued 21\,cm line emission in HIPASS
  (\citealt{Koribalski:2004, Meyer:2004,
    Wong:2006}).}\label{figure:luminosity_class}
\end{figure}

\begin{figure}
\vspace{6pt}
\centering
\includegraphics[width = 1.0\columnwidth]{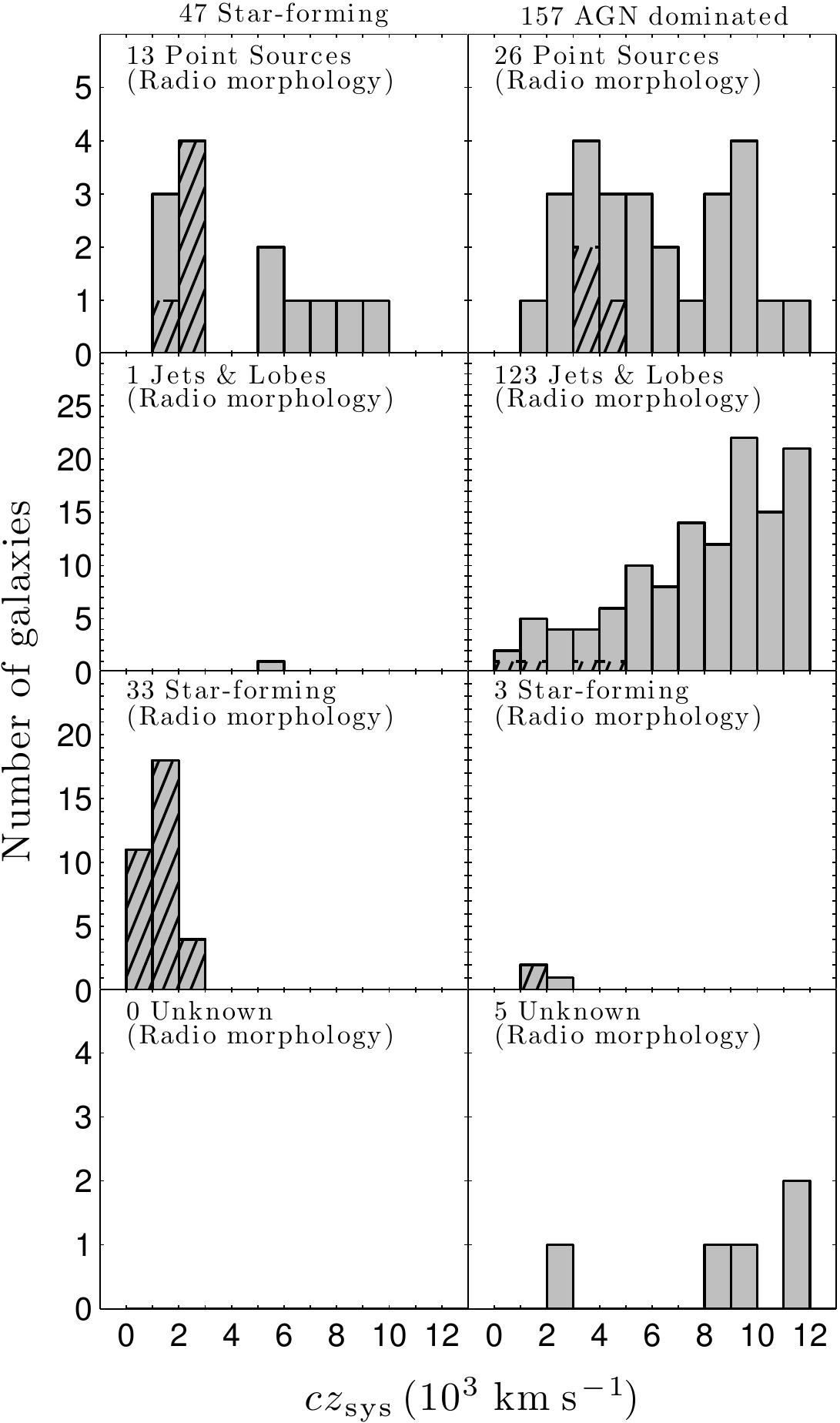}
\caption{As Fig.\,\ref{figure:luminosity_class}, but for the
  distribution of redshifts.}\label{figure:redshift_class}
\end{figure}

\subsection{Comparison with the 2\,Jy
  sample}\label{section:2jy_sample}

\cite{Morganti:2001} used the ATCA, the VLA and the Westerbork
Synthesis Radio Telescope to search for \mbox{H\,{\sc i}} absorption
in 23 radio galaxies (at $z<0.22$ and $\delta < +10\degr$) selected
from the 2\,Jy sample \citep{Wall:1985}. We can use this relatively
homogeneous set of observations to determine whether our
non-detections in HIPASS are consistent with what we would expect from
existing detections of absorption. In five of these radio galaxies,
Morganti et al. detected \mbox{H\,{\sc i}} absorption, of which
NGC\,5090 ($cz = 3421\pm21$\,km\,s$^{-1}$) and 3C\,353 ($cz =
9120\pm59$\,km\,s$^{-1}$) are within the volume surveyed by
HIPASS. The 21\,cm spectra of these two galaxies exhibit peak
absorption of 8 and 10\,mJy, with FWHMs of approximately 100 and
200\,km\,s$^{-1}$, respectively. Given the noise and baseline ripple
confusion in the HIPASS spectra, our non-detection of \mbox{H\,{\sc
    i}} absorption in these radio galaxies is consistent with the
expected strength of these lines.

\subsection{The HIPASS detection rate}

We obtain detection rates for associated absorption in HIPASS of
2.0\,per\,cent (4/204) for the total sample, 4.4\,per\,cent (4/90) for
Sample 1 and 1.6\,per\,cent (3/189) for Sample 2. While such a small
number of detections does not allow us to draw strong conclusions
about the population, we can attempt to understand these rates in the
context of the HIPASS survey parameters and the properties of
individual galaxies in the sample.

In Fig.\,\ref{figure:luminosity}, we show the 1.4\,GHz radio power
versus systemic redshift (for all 204 galaxies in our sample), and the
far-infrared luminosity for those 86 galaxies that either have a
detection at 60 or 100\,$\mathrm{\mu}$m in the \emph{Infrared
  Astronomical Satellite} (\emph{IRAS}) Faint Source, Point Source and
Galaxy Catalogues \citep{Beichman:1988, Rice:1988, Knapp:1989,
  Moshir:1992, Sanders:2003}. The radio power is estimated using the
larger of either the 843\,MHz SUMSS (assuming a spectral index of
$-0.6$) or 1.4\,GHz NVSS total flux densities, and thereby accounting
for components that might be present in SUMSS but missing in the NVSS
images.

We calculate the far-infrared luminosity using an estimate of the flux
density ($S_\mathrm{FIR}$) between 42.5 and 122.5\,$\mathrm{\mu}$m,
which is given by \citep{Helou:1985}
\begin{equation}
  S_\mathrm{FIR} = 1.26 \times 10^{-14}\,(2.58\,S_\mathrm{60\mu m} + S_\mathrm{100\mu m})\,\mathrm{W\,m^{-2}},
\end{equation}
where $S_\mathrm{60\mu m}$ and $S_\mathrm{100\mu m}$ are the 60 and
100\,$\mu$m flux densities in units of Jy. For those galaxies where
measurements of only $S_\mathrm{60\mu m}$ or $S_\mathrm{100\mu m}$ are
available, we use $\langle\log_{10}(S_\mathrm{100\mu
  m}/S_\mathrm{60\mu m})\rangle = 0.3$, which is the average
calculated from the \emph{IRAS} Bright Galaxy sample by
\cite{Soifer:1989}. Galaxies that are identified as star forming
exhibit a strong correlation between their radio and far-infrared
luminosities \citep[e.g][]{Helou:1985, Devereux:1989,
  Condon:1991}. For a large sample of spectroscopically identified
star-forming galaxies, \cite{Mauch:2007} measured this relationship to
be
\begin{equation}
  \log_{10}(P_{1.4}) = (1.06 \pm 0.01) \log_{10}(L_{\mathrm{FIR}}) + (11.1 \pm 0.1),
\end{equation}
with a maximum deviation in radio power of approximately one
decade. We use this to identify galaxies in our sample that are
star-forming, classifying AGN dominated galaxies as those that do not
follow this relationship or do not have detections in both the 60 and
100\,$\mathrm{\mu}$m bands.

Based on the $P_{1.4}$--$L_\mathrm{FIR}$ relation, we estimate that
20\,per\,cent (47/204) of our galaxies are star forming, which is
consistent with the 42 predicted using the local radio luminosity
function of \cite{Mauch:2007}. Considering the relative radio and
near-infrared morphologies of these star-forming galaxies, 13 are
unresolved point sources, 33 have extended emission that is consistent
with star formation, and only 1 is identified as having jets and
lobes\footnote{2MASX\,J18324117-3411274, which on visual inspection of
  the SUMSS image could also be confused with nearby point source
  companions.}. For the remaining 157 galaxies in our sample, which we
classify as AGN dominated, 26 are unresolved point sources, 123 are
identified as having jets and lobes, 3 have extended emission
consistent with star formation\footnote{NGC\,2559, WKK\,4748 and
  NGC\,5903, all of which were not detected in either of the 60 and
  100\,$\mu$m bands of \emph{IRAS}.}, and 5 have unknown
structure. Hence, there is a clear consistency between the
classifications based on the radio and far-infrared luminosities and
the radio and near-infrared morphologies. We summarize these
classifications in Figs\,\ref{figure:luminosity_class} and
\ref{figure:redshift_class}, showing their distribution as a function
of 1.4\,GHz radio power and redshift.

For the sub-sample of star-forming galaxies, two factors significantly
reduce the likelihood of detecting \mbox{H\,{\sc i}} absorption in
HIPASS -- the predominance of \mbox{H\,{\sc i}} emission (which arises
from the large reservoirs of gas required to form stars) and the
distribution of the continuum flux density over the extended stellar
disc. In the case of the former, the spatial distribution of the
emission is typically unresolved by HIPASS and so acts to
significantly mask any potential absorption of the background
continuum at low redshifts. Furthermore, the continuum emission is
extended over kpc scales, effectively reducing the covering factor $f$
and hence the likelihood of detecting absorption against a small
fraction of the total flux density. If we assume that the sizes of
cold and dense \mbox{H\,{\sc i}} gas clouds are typically 100\,pc
\citep[e.g][]{Braun:2012, Curran:2013b}, then the fraction of radio
emission obscured by a single absorbing cloud will be $f \sim 0.01$,
which is equivalent to $S_{1.4}\sim3$\,mJy for the flux density limit
of our sample, and so well below the noise level. While 33 of the
star-forming galaxies are identified morphologically as having
extended emission, some of the more compact point sources will have
star formation concentrated within the sub-kpc nuclear region,
effectively increasing the likelihood of absorption detection. This is
certainly the case for the single detection of \mbox{H\,{\sc i}}
absorption we obtain in our sub-sample of star-forming galaxies,
Arp\,220, where a significant fraction of the nuclear radio emission
is thought to be generated by starburst activity ($\sim
240$\,M$_{\odot}$\,yr$^{-1}$; \citealt{Anantharamaiah:2000}).

A far smaller fraction of the AGN-dominated radio galaxies have
\mbox{H\,{\sc i}} emission detected in HIPASS compared with those that
are star forming. This is in part due to their distribution towards
higher redshifts, but also that many of these sources will be hosted
by massive, neutral gas-poor, early-type galaxies
(e.g. \citealt{Bregman:1992}). However, while fewer have emission
lines that could mask the detection of absorption at low redshifts,
the majority have morphologies (jets and lobes) that are extended over
scales greater than 45\,arcsec (the typical spatial resolution of both
NVSS and SUMSS), which at the median redshift of $cz =
6000$\,km\,s$^{-1}$ equates to physical scales greater than $\sim
20$\,kpc. The likelihood of absorption against these extended sources
is low since most of the continuum emission will not be obscured by
the discs or rings in which we expect the absorbing \mbox{H\,{\sc i}}
gas to be located. In the special case of Centaurus\,A, the proximity
of this radio galaxy to the Milky Way means that, while only a
fraction of the total continuum emission is concentrated within the
nucleus of the galaxy, we can still detect significant absorption of
$\Delta{S} \approx 1$\,Jy against the core. Furthermore, the
\mbox{H\,{\sc i}} emission and absorption are spatially resolved and
so can be identified as separate components. If Centaurus\,A were
instead located at the sample median redshift of $cz =
6000$\,km\,s$^{-1}$, both the emission and absorption (which would
decrease to less than 13\,mJy) would no longer be detectable with
HIPASS.

Our two remaining detections of absorption occur in AGN-dominated
radio galaxies, 2MASX\,J13084201-2422581 and NGC\,5793, both with
continuum radio emission that is compact with respect to the size of
the stellar disc of the galaxy (see
Fig.\,\ref{figure:image_overlays_cabb}). As we have already noted in
Section\,\ref{section:individual_sources}, these galaxies are both
classified as having edge-on disc morphologies (with inclinations of
$i \approx 75\degr$) and Seyfert~2 AGN activity. It is towards these
compact radio galaxies that we would expect to have the highest
detection rate for absorption, where a significant fraction of the
total continuum flux density will be absorbed by the chance alignment
of foreground cold and dense \mbox{H\,{\sc i}} gas. At higher
redshifts, where the 21\,cm emission line is not easily detectable in
HIPASS, the compact and radio-loud nuclear starbursts will also
contribute significantly to the detection rate, as was seen for
Arp\,220. If we consider just the point sources that do not have
catalogued 21\,cm line emission, then we obtain a detection rate for
absorption of 6\,per\,cent (2/31), which is approaching the typical
rates obtained by targeted searches of compact radio sources (see
e.g. \citealt{Allison:2012a} and references therein).

\subsection{Comparison with the ALFALFA pilot
  survey}\label{section:alfalfa_survey}

The ALFALFA survey on the Arecibo Telescope \citep{Giovanelli:2005} is
the only other existing large field-of-view survey for \mbox{H\,{\sc
    i}} gas in the local Universe, which when completed will map
$7000\,\deg^{2}$ of the sky in the redshift range $-2000 < cz <
19\,000\,\mathrm{km}\,\mathrm{s}^{-1}$. \citet{Darling:2011} recently
conducted a blind pilot survey of \mbox{H\,{\sc i}} absorption in the
volume bounded by $-650 < cz < 17\,500\,\mathrm{km}\,\mathrm{s}^{-1}$
and $517\,\deg^{2}$ (1.3\,per\,cent of the celestial sphere and
7.4\,per\,cent of the full ALFALFA footprint). They found no
intervening absorbers (which is consistent with the redshift search
path and column density limits) and a single strong absorption line
($\tau \approx 0.64$) at $cz \approx
10\,800\,\mathrm{km}\,\mathrm{s}^{-1}$, associated with the
interacting luminous infrared galaxy UGC\,6081 that had previously
been detected by \citet{Bothun:1983} and \citet{Williams:1983}.

To compare their result with our HIPASS search, we again use the local
radio luminosity function of \cite{Mauch:2007} to estimate the
expected number of galaxies above a flux density limit of $S_{1.4} =
42$\,mJy (defined by a $5\,\sigma$ detection of
absorption\footnote{The per-channel noise in the smoothed ALFALFA
  spectra is 2.2\,mJy.} with an optical depth of $\tau > 0.3$), within
the comoving volume bounded by 517\,deg$^{2}$ of sky and the redshift
range $0 < cz < 17\,500\,\mathrm{km}\,\mathrm{s}^{-1}$ (approximately
$8 \times 10^{-4}$\,Gpc$^{3}$). This yields approximately 29 galaxies,
of which 19 are AGN dominated and 10 are star forming. The total
detection rate based on this sample is therefore $\sim 3$\,per\,cent
(1/29), which is consistent with our results. We note that
10\,per\,cent of the redshift range was found to be unusable by
\cite{Darling:2011}, due to contamination from RFI and Galactic 21\,cm
emission, and so the expected detection rate is in fact slightly
higher. UGC\,6081 is not in the region of sky observed by \emph{IRAS},
and so we cannot use the far-infrared versus radio luminosity
relationship to classify this galaxy. The galaxy is in the process of
a merger, exhibiting two radio nuclei that are separated by only
16\,arcsec in the Faint Images of the Radio Sky at Twenty Centimeters
survey (\citealt{Becker:1995}), indicating that the radio emission may
be arising from nuclear starburst activity in a similar mode to
Arp\,220. The radio power (assuming a total flux density of $S_{1.4} =
170$\,mJy; \citealt{White:1997}) at the redshift of the galaxy is
$P_{1.4} \approx 5 \times 10^{23}$\,W\,Hz$^{-1}$, which is consistent
with either a highly luminous starburst or AGN activity. UGC\,6081
would likely be classified as either a compact AGN or nuclear
starburst in our sample, which is consistent with the majority of our
detections (excluding Centaurus\,A as a special case).

\subsection{Implications for future H\,{\sevensize\bf I} absorption
    surveys}

\begin{figure}
\centering
\includegraphics[width = 1.0\columnwidth]{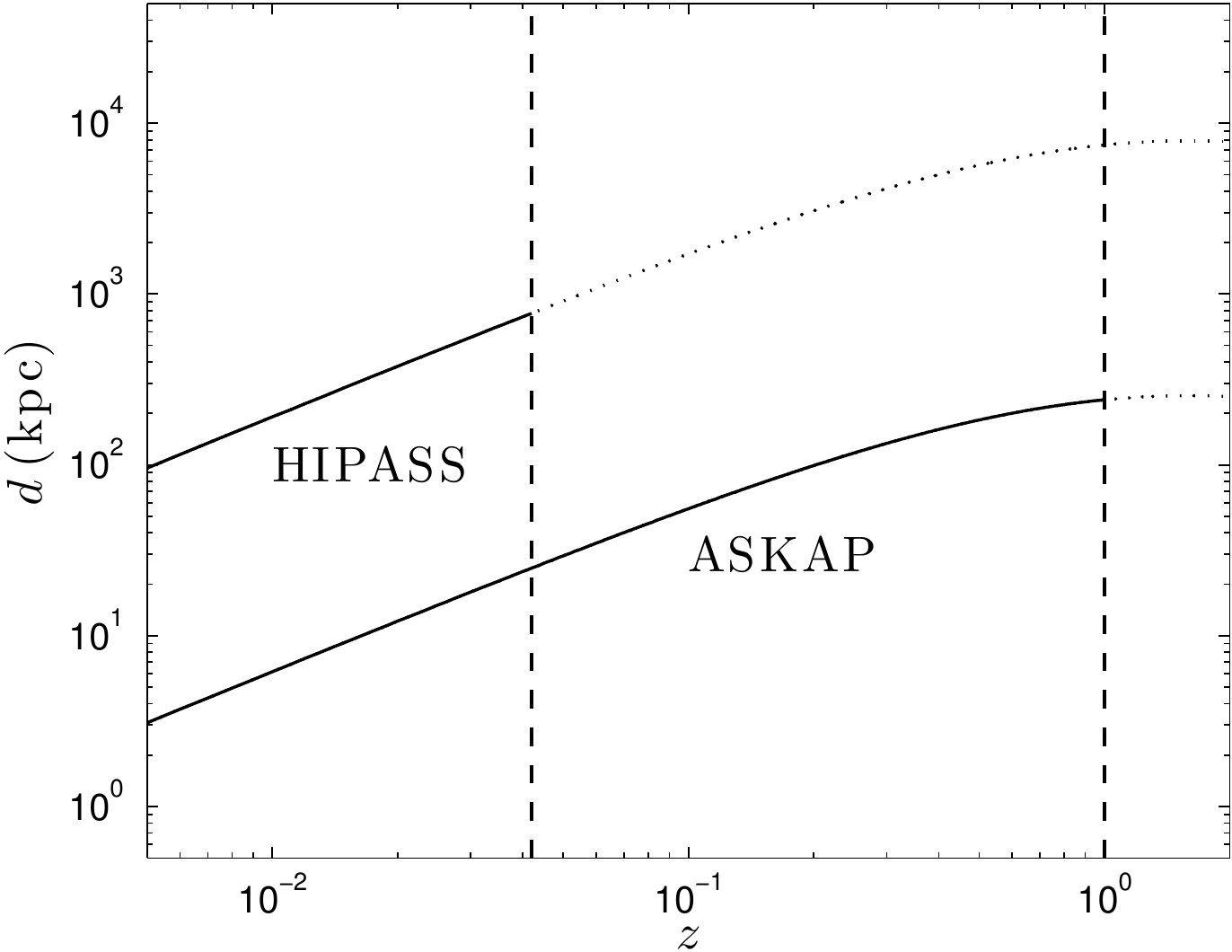}
\caption{Physical size versus redshift for the 15.5\,arcmin HIPASS
  beam and the upper-limit 30\,arcsec beam expected for a future
  \mbox{H\,{\sc i}} absorption survey using ASKAP. The solid lines
  indicate the physical scale of the beam within the redshift range
  indicated by the vertical dashed lines. At the higher redshifts
  available to an \mbox{H\,{\sc i}} absorption survey using ASKAP, a
  30\,arcsec beam will be sensitive to the physical scales resolved by
  HIPASS.}\label{figure:size_redshift}
\end{figure}

We can use our results to estimate the number detections that might be
achievable with the full ALFALFA survey. Considering a blind survey
for associated absorption, conducted over 7000\,deg$^{2}$ of the sky
in the redshift range $0 < cz < 19\,000$\,km$^{-1}$ (equating to a
comoving volume of 0.0136\,Gpc$^{3}$), we use the local radio
luminosity function of \cite{Mauch:2007} to predict a total of 455
galaxies above a detection flux limit of 42\,mJy (see
Section\,\ref{section:alfalfa_survey} for an explanation of this
limit). By simply applying our total detection rate from HIPASS, we
predict approximately 10 detections of associated absorption, while
applying the rate estimated for the pilot survey of
\cite{Darling:2011} yields approximately 16 (which is driven by the
fractional increase in volume of the full survey). Since the ALFALFA
survey probes higher redshifts than HIPASS, we expect that the
detection rate amongst a radio flux density selected sample of
galaxies will be higher, due to a decrease in the fraction of diffuse
star-forming galaxies. We therefore predict a factor of 3--4 increase
in the number of detections of associated absorption compared with
what we have achieved in HIPASS.

Our results have positive implications for proposed all-sky absorption
surveys on the precursor telescopes to the SKA, which will be able to
probe the \mbox{H\,{\sc i}} content of the Universe up to $z \approx
1$. In Fig.\,\ref{figure:size_redshift}, we show that the physical
scales to which HIPASS is sensitive in the local Universe are well
matched at higher redshifts to the smaller beam sizes of the proposed
First Large Absorption Survey in \mbox{H\,{\sc i}} on the Australian
Square Kilometre Array Pathfinder (ASKAP; \citealt{Deboer:2009}). The
HIPASS detections show that it will be possible to use ASKAP (with a
spatial resolution of $d \sim 100$\,kpc at $z \sim 0.5$) to detect
strong absorption systems associated with compact AGN and nuclear
starbursts at redshifts in the range $0.5 < z < 1.0$, probing an epoch
of the Universe not yet explored by all-sky \mbox{H\,{\sc i}} surveys.

Furthermore, our results indicate that such strong absorption systems
detected in future all-sky surveys may well provide excellent targets
for luminous H$_{2}$O megamaser detection at redshifts greater than
0.1, where so few have been discovered\footnote{SDSS J0804+3607 at $z
  = 0.66$ \citep{Barvainis:2005} and 4C\,+05.19 at $z = 2.64$
  \citep{Impellizzeri:2008}, towards which multiple intervening and
  associated \mbox{H\,{\sc i}} absorption lines have been detected
  \citep{Moore:1999, Curran:2007b, Curran:2011c,
    Tanna:2013}.}. \citet{Taylor:2002} showed that for the sample of
known H$_{2}$O megamaser galaxies at that time, the detection rate of
\mbox{H\,{\sc i}} absorption was greater than 42\,per\,cent, higher
than the typical rates achieved with targeted surveys of compact radio
sources (see \citealt{Allison:2012a} and references therein). Of the
Seyfert~2 galaxies in which we have detected absorption, NGC\,5793 has
known H$_{2}$O megamaser emission from within a sub-pc disc around the
AGN \citep{Hagiwara:1997}, while 2MASX\,J13084201-2422581, although
not detected, has been a candidate for several large-scale megamaser
surveys (e.g. \citealt{Braatz:1996, Sato:2005, Kondratko:2006}). VLBI
studies at high spatial resolution show that the optically thick
\mbox{H\,{\sc i}} absorption in these galaxies probably arises within
an edge-on disc on 100\,pc scales. If dense molecular gas also exists
on sub-pc scales, it is likely to be distributed as a circumnuclear
disc that is similarly orientated and so generate significant
amplification of H$_{2}$O emission towards us, thereby providing
favourable conditions for the detection of megamasers at cosmological
redshifts. We also note that radio-loud nuclear starburst galaxies
such as Arp\,220, which is host to multiple regions of OH megamaser
emission within its double nuclei \citep{Lonsdale:1998, Rovilos:2003},
are likely to produce strong 1.6\,GHz emission that can be detected in
the wide frequency bands of these future \mbox{H\,{\sc i}} surveys.

\section{Summary}

We have used archival data from HIPASS to search for \mbox{H\,{\sc i}}
21\,cm absorption within a sample of 204 nearby radio and star-forming
galaxies, achieving a total detection rate of 2.0\,per\,cent
(4/204). Of these detections, three are found against compact radio
sources (two AGN and a nuclear starburst), while the fourth is within
the nearby large radio galaxy Centaurus\,A, which would not have been
seen at larger redshifts. Although susceptible to low number
statistics, the detection rate against just the morphologically
compact radio sources (both AGN and nuclear starbursts) is higher than
the total rate and closer to the typical values obtained from targeted
surveys of compact sources.

In the case of 2MASX\,J13084201-2422581, the absorption line had not
been previously detected in the literature, highlighting the
serendipitous advantages of performing such a low-sensitivity all-sky
survey for absorption. The 21\,cm line profile is similar to that seen
in other edge-on Seyfert~2 galaxies, indicating that the absorption
may arise within a disc of \mbox{H\,{\sc i}} gas on 100\,pc scales. A
follow-up observation with the ATCA at 10\,arcsec spatial resolution
demonstrates that all of the absorption detected in the HIPASS
spectrum arises against the compact radio nucleus of this galaxy. The
higher sensitivity and spectral resolution of the CABB system reveals
the presence of a second blueshifted component that might signify a
200\,km\,s$^{-1}$ outflow of neutral gas.

The detection rate we achieve with HIPASS is consistent with that
found for the ALFALFA pilot survey carried out by
\cite{Darling:2011}. We predict that the full ALFALFA survey will
yield three to four times as many associated absorption systems as we
have achieved with HIPASS, and that future all-sky absorption surveys
at higher redshifts should yield many more new detections. HIPASS is
sensitive to only the strongest absorption lines, which appear to be
dominated by galaxies that exhibit edge-on discs of atomic gas and
high columns of nuclear molecular gas that exhibit H$_{2}$O megamaser
emission. We predict that such systems detected in future all-sky
surveys have the potential to provide excellent targets for the
detection of luminous H$_{2}$O megamaser emission close to the AGN,
with the potential for direct measurement of black hole masses at
cosmological redshifts \citep{Miyoshi:1995, Kuo:2011} and independent
determination of the Hubble Constant \citep{Reid:2013}.

\section*{Acknowledgements} 

We thank B\"{a}rbel Koribalski, Peter Tuthill and Geraint Lewis for
useful discussions, and William Wilhelm and Stephen Curran for their
help with querying data bases. We also thank the anonymous referee for
useful comments that helped improve this paper. JRA acknowledges
support from an ARC Super Science Fellowship. Parts of this research
were conducted by the Australian Research Council Centre of Excellence
for All-sky Astrophysics (CAASTRO), through project number
CE110001020. The Parkes telescope and ATCA are part of the Australia
Telescope which is funded by the Commonwealth of Australia for
operation as a National Facility managed by CSIRO. Computing
facilities were provided by the High Performance Computing Facility at
the University of Sydney. This research has made use of the NASA/IPAC
Extragalactic Database (NED) which is operated by the Jet Propulsion
Laboratory, California Institute of Technology, under contract with
the National Aeronautics and Space Administration; NASA's Astrophysics
Data System Bibliographic Services; the SIMBAD data base and VizieR
catalogue access tool, both operated at CDS, Strasbourg, France.

\bibliographystyle{mn2e.bst}
\bibliography{./bibliography.bib}

\appendix

\section{Radio -- optical source matches excluded from Sample
  1}\label{section:sample1_exclusions}

\noindent \emph{PKS\,0056-572} (SUMSS\,J005846-565912, $S_{843} =
479$\,mJy) is reported by \cite{Johnston:1995} as a quasar with a
redshift of $z = 0.018$; however, there is no apparent reference to
the origin of this measurement and so we have excluded this object
from our sample.

\noindent \emph{Fairall\,0216} (SUMSS\,J005825-664841, $S_{843} =
382$\,mJy) is a galaxy pair with a redshift reported by
\cite{Fairall:1980} of $z \approx 0.033$. However, \cite{Garilli:1993}
and \cite{Dalton:1994} report a significantly higher redshift for one
of the component galaxies ($z \sim 0.07$) and no redshift is given for
the other. Given the potential unreliability of the original redshift
measurement and confusion as to which component the radio source
belongs, we exclude this object from our sample.

\noindent \emph{UM\,310} (NVSS\,J011517-012705, $S_{1.4} = 1080$\,mJy)
is reported in the 6dF\,GS catalogue to have a redshift of $z =
0.014\,36$. However, the 6dF spectral features are weak and this
estimate is inconsistent with the much higher value of $z = 1.36$
measured by \cite{Lewis:1979} based on the \mbox{C\,{\sc iv}}
$\lambda$1549 emission line. Since there is a high likelihood that
this object is located outside of the HIPASS volume, we do not include
it in our sample.

\noindent \emph{PKS\,0131-522} (SUMSS\,J013305-520005, $S_{843} =
366$\,mJy) is reported by \cite{Johnston:1995} as a quasar with a
redshift of $z = 0.02$; however, there is no apparent reference to the
origin of this measurement and so we have excluded this object from
our sample.

\noindent \emph{PKS\,0208-512} (SUMSS\,J021046-510102, $S_{843} =
3370$\,mJy) has a redshift of $z = 0.031\,56$ in the 6dF\,GS
catalogue. However, this estimate is inconsistent with the
significantly higher value of $z = 1.003$ recorded in the quasar
catalogue by \citealt{Veron-Cetty:2010}, and so we have excluded this
object from our sample.

\noindent \emph{NGC\,1310} (NVSS\,J032102-370604, $S_{1.4} =
457$\,mJy) is a spiral galaxy in the Fornax Cluster with a
redshift of $0.006$ \citep{Ferguson:1989, Drinkwater:2001}. Its close
proximity to bright extended radio emission from Fornax A, which is
apparent upon visual inspection of SUMSS images, means that the
integrated flux density recorded in the NVSS catalogue is likely to be
unreliable. We have therefore excluded this object from our sample.

\noindent \emph{SNR\,J050854-684447} (SUMSS\,J050859-684333, $S_{843}
= 653$\,mJy) is a flat-spectrum supernova remnant located within the
Large Magellanic Cloud \citep{Mathewson:1973}.

\noindent \emph{6dF\,J0653599-415145} (SUMSS\,J065400-415144, $S_{843}
= 728$\,mJy) has a measured redshift of $z = 0.00$ in the 6dF\,GS
catalogue. However, \cite{Sadler:2014} find that the 6dF spectrum is
contaminated by foreground Galactic nebular emission lines and their
re-measured value of $z = 0.0908$ puts the host galaxy well outside of
the HIPASS volume.

\noindent \emph{PKS\,0959-443} (MGPS-2\,J100200-443806, $S_{843} =
1790$\,mJy) was originally measured by \cite{Wright:1977} to have a
redshift of $z = 0.840$ based on quasar emission lines. After
obtaining higher signal-to-noise data, \cite{Wright:1979} found that
the identification of the emission lines was incorrect and used the
strong galaxy-type absorption system to measure a much lower redshift
of $z = 0.021$, re-classifying this object as a very compact nearby
galaxy. Given the inconsistency of these redshift estimates, and that
this source also lies in the Galactic plane, we have decided to
exclude this from our sample. It should be noted that no 21\,cm
absorption line is seen towards this source either in the HIPASS data
or in the literature.

\noindent \emph{GQ\,1042+0747} (NVSS\,J104257+074850, $S_{1.4} =
383$\,mJy) was incorrectly matched with the background radio
source. This dwarf spiral galaxy ($z = 0.033\,21$) forms a close pair
with the background quasar SDSS\,J104257.58+074850.5 ($z =
2.665\,21$). Using 21\,cm spectroscopy of the radio source, with the
Green Bank Telescope (GBT), VLA and Very Long Baseline Array,
\cite{Borthakur:2010} detected intervening \mbox{H\,\sc{i}} absorption
in the foreground galaxy at $z = 0.03310$. The GBT spectrum exhibits
an absorption line with an FWHM of 3.6\,km\,s$^{-1}$ and peak flux
density of $\sim18$\,mJy, and is therefore consistent with this line
not being detected in the HIPASS spectrum.

\noindent \emph{PKS\,1221-42} (SUMSS\,J122343-423531, $S_{843} =
3110$\,mJy) was reported by \cite{Sadler:2014} to have an incorrectly
measured redshift of $z = 0.0266$ in the 6dF\,GS catalogue, and that
the correct value is $z = 0.1706$ \citep{Simpson:1993,
  Johnston:2005}. Since this is outside of the HIPASS volume we have
excluded the source from our sample.

\noindent \emph{PKS\,1510-08} (NVSS\,J151250-090600, $S_{1.4} =
2700$\,mJy) is a quasar with a low redshift measurement of $z =
0.006818$ reported in the 6dF\,GS catalogue. However, this estimate is
inconsistent with the much higher value of $z = 0.360$ measured by
\cite{Thompson:1990}, and so we have excluded this source from our
sample.

\noindent \emph{NGC\,6210} (NVSS\,J164429+234759, $S_{1.4} =
298$\,mJy) is a planetary nebula ($cz =
-36\,\mathrm{km}\,\mathrm{s}^{-1}$; \citealt{Wilson:1953}).

\noindent \emph{6dF\,J1716205-615706} (SUMSS\,J171620-615707,
$S_{843} = 258$\,mJy) is identified as a star in the 6dF\,GS catalogue
($cz = -57\,\mathrm{km}\,\mathrm{s}^{-1}$; \citealt{Jones:2009}).

\noindent \emph{PKS\,1935-526} (SUMSS\,J193953-523039, $S_{843} =
460$\,mJy) is identified by \cite{Grandi:1983} as a dwarf galaxy with
a single redshift measurement of $z = 0.03$. Since there has been no
further spectroscopic verification of this redshift measurement, we
have excluded this source from our sample.

\newpage
\onecolumn

\begin{landscape}

\section{Galaxies searched for associated H\,{\sevensize\bf I}
  absorption}\label{section:full_sample}

\setlength\tabcolsep{4pt}
\begin{longtable}{lccrrlrrrrrrrrrrrrr}
  \caption{Properties of 204 galaxies searched for \mbox{H\,{\sc i}}
    absorption in HIPASS. Right ascension and declination are given
    for the centre position used to extract the HIPASS
    spectrum. $cz_{\rm sys}$ and $\sigma_{cz}$ are the mean and
    uncertainty in the systemic redshift. $S_{843}$ and
    $\sigma_\mathrm{843}$ are the mean and uncertainty in the
    SUMSS/MGPS-2 flux density at 843\,MHz
    \citep{Mauch:2003,Murphy:2007}. $S_{1.4}$ and
    $\sigma_\mathrm{1.4}$ are the mean and uncertainty in the NVSS
    total flux density at 1.4\,GHz \citep{Condon:1998}. $n_{843}$ and
    $n_{1.4}$ are the number of radio components in the NVSS and
    SUMSS/MGPS-2 images, respectively \citep{VanVelzen:2012}. Note
    that where available we have used the SUMSS/MGPS-2 and NVSS total
    flux densities compiled by \citet{VanVelzen:2012}. $S_{\rm CHI}$
    is the beam-weighted flux density extracted within a single HIPASS
    beamwidth from the 1.4\,GHz CHIPASS compact source map
    (\citealt{Calabretta:2014}). $S_{100\,\mu\mathrm{m}}$ and
    $S_{60\,\mu\mathrm{m}}$ are the 100 and 60\,$\mu$m flux densities
    from the IRAS Faint Source, Point Source and Galaxy Catalogues
    \citep{Beichman:1988, Rice:1988, Knapp:1989, Moshir:1992,
      Sanders:2003}. $K_{s}$ is the apparent $K_{s}$-band magnitude,
    within the 20\,mag\,arcsec$^{-2}$ isophote, from the 2MASS
    (\citealt{Skrutskie:2006}). $S_{\rm HI}$ is the integrated
    \mbox{H\,{\sc i}} emission from HIPASS
    (\citealt{Koribalski:2004,Meyer:2004,Wong:2006}). The radio
    morphology classifications are as those given by
    \citet{VanVelzen:2012}, where p = point sources, g = star-forming
    galaxies, j = jets and lobes and u = unknown. The full version of
    this table is available
    online.}\label{table:full_sample}\\
  \hline
  \multicolumn{1}{l}{Name} & \multicolumn{1}{c}{RA} & \multicolumn{1}{c}{Dec.} & \multicolumn{1}{c}{$cz_{\rm sys}$} & \multicolumn{1}{c}{$\sigma_{cz}$} & \multicolumn{1}{l}{$z$ ref.} & \multicolumn{1}{c}{$S_{843}$} & \multicolumn{1}{c}{$\sigma_{843}$} & \multicolumn{1}{c}{$n_{843}$} & \multicolumn{1}{c}{$S_{1.4}$} & \multicolumn{1}{c}{$\sigma_{1.4}$} & \multicolumn{1}{c}{$n_{1.4}$} & \multicolumn{1}{c}{$S_{\rm CHI}$} & \multicolumn{1}{c}{$S_{100\,\mu\mathrm{m}}$} & \multicolumn{1}{c}{$S_{60\,\mu\mathrm{m}}$} & \multicolumn{1}{c}{$K_{s}$} & \multicolumn{1}{c}{$S_{\rm HI}$} & \multicolumn{1}{r}{Mor.} & \multicolumn{1}{r}{Sam.} \\
  & \multicolumn{2}{c}{(J2000)} & \multicolumn{2}{c}{(km\,s$^{-1}$)} & & \multicolumn{1}{c}{(mJy)} & \multicolumn{1}{c}{(mJy)} & & \multicolumn{1}{c}{(mJy)} & \multicolumn{1}{c}{(mJy)} & & \multicolumn{1}{c}{(mJy)} & \multicolumn{1}{c}{(mJy)} & \multicolumn{1}{c}{(mJy)} & \multicolumn{1}{c}{(mag)} & \multicolumn{1}{c}{(Jy\,km\,s$^{-1}$)} & & \\
  \hline
  PKS\,0000-550 & 00 03 10.63 & -54 44 56.1 & 9767 & 24 & \citet{Huchra:2012} & 1549 & 55 & 1 & $-$ & $-$ & $-$ & 1050 & $-$ & $-$ & 10.30 & $-$ & j & 1,2\\
  ESO\,193-G017 & 00 05 07.95 & -50 49 18.1 & 11\,169 & 45 & \citet{Dressler:1988} & 281 & 10 & 1 & $-$ & $-$ & $-$ & 93 & $-$ & $-$ & 11.00 & $-$ & j & 1\\
  IC\,1531 & 00 09 35.54 & -32 16 36.7 & 7632 & 45 & \citet{Jones:2009} & 559 & 17 & 1 & 433 & 13 & 2 & 642 & $-$ & $-$ & 9.71 & $-$ & j & 1,2\\
  NGC\,0025 & 00 09 59.28 & -57 01 15.0 & 9465 & 32 & \citet{Lauberts:1989} & 619 & 24 & 1 & $-$ & $-$ & $-$ & 445 & $-$ & $-$ & 9.98 & $-$ & j & 1,2\\
  NGC\,0055 & 00 14 53.60 & -39 11 47.9 & 129 & 2 & \citet{Koribalski:2004} & $-$ & $-$ & $-$ & 225 & 8 & 4 & 417 & 174\,000 & 77\,000 & 6.56 & 1990 & g & 2\\
  NGC\,0092 & 00 21 31.69 & -48 37 29.3 & 3219 & 7 & \citet{Meyer:2004} & 252 & 8 & 1 & $-$ & $-$ & $-$ & 365 & $-$ & $-$ & 9.43 & 12 & p & 1\\
  NGC\,0193 & 00 39 18.58 & +03 19 52.7 & 4414 & 15 & \citet{Ogando:2008} & $-$ & $-$ & $-$ & 1277 & 32 & 2 & 1710 & $-$ & $-$ & 9.34 & $-$ & j & 2\\
  NGC\,0215 & 00 40 48.87 & -56 12 50.7 & 8154 & 45 & \citet{Jones:2009} & 310 & 10 & 2 & $-$ & $-$ & $-$ & 334 & $-$ & $-$ & 9.96 & $-$ & j & 2\\
  NGC\,0253 & 00 47 33.13 & -25 17 19.7 & 243 & 2 & \citet{Koribalski:2004} & $-$ & $-$ & $-$ & 5572 & 121 & 8 & 5630 & 1290\,000 & 968\,000 & 3.82 & 694 & g & 1,2\\
  UGC\,00583 & 00 56 25.59 & -01 15 44.9 & 11\,461 & 3 & \citet{Aihara:2011} & $-$ & $-$ & $-$ & 1370 & 26 & 4 & 3240 & $-$ & $-$ & 10.60 & $-$ & j & 2\\
  NGC\,0349 & 01 01 50.75 & -06 47 59.3 & 5986 & 18 & \citet{Ogando:2008} & $-$ & $-$ & $-$ & 296 & 8 & 9 & 242 & $-$ & $-$ & 9.32 & $-$ & j & 2\\
  IC\,1623B & 01 07 47.56 & -17 30 25.1 & 6087 & 45 & \citet{Jones:2009} & $-$ & $-$ & $-$ & 248 & 10 & 1 & 217 & $-$ & $-$ & 10.00 & $-$ & p & 2\\
  \hline
\end{longtable}
\end{landscape}

\begin{figure*}
\centering
\includegraphics[width = 1.0\textwidth]{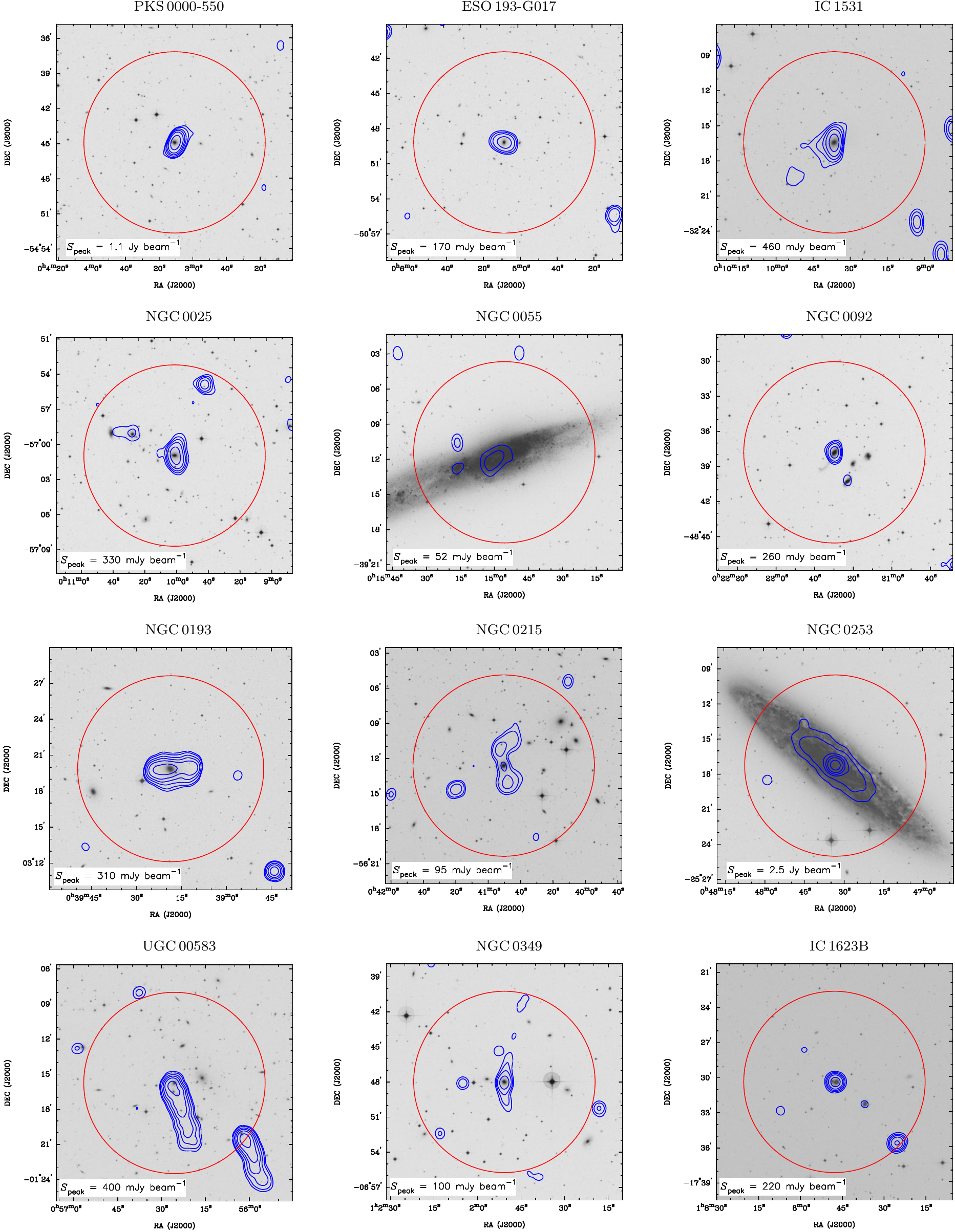}
\caption{Images of 204 galaxies searched for \mbox{H\,{\sc i}}
  absorption using HIPASS. The blue contours denote the 1, 2, 5, 10,
  20, 50\,per\,cent levels of the peak radio flux density within the
  HIPASS beamwidth, from either NVSS ($\delta > -30\degr$, $\nu =
  1.4$\,GHz, typical beam FWHM = 45\,arcsec) or SUMSS/MGPS-2 ($\delta
  < -30\degr$, $\nu = 843$\,MHz, typical beam FWHM = 45\,arcsec). For
  clarity, we exclude those radio contours that are less than five
  times the survey rms. The large red circle represents the gridded
  HIPASS beamwidth of 15.5\,arcmin. The grey-scale images represent
  optical $B_{j}$-band photometry from the SuperCosmos Sky Survey,
  using the UK Schmidt and Palomar Oschin Schmidt telescopes
  \citep{Hambly:2001}. The full version of this figure is available
  online.}\label{figure:image_overlays_total}
\end{figure*}

\begin{figure*}
\centering
\includegraphics[width = 1.0\textwidth]{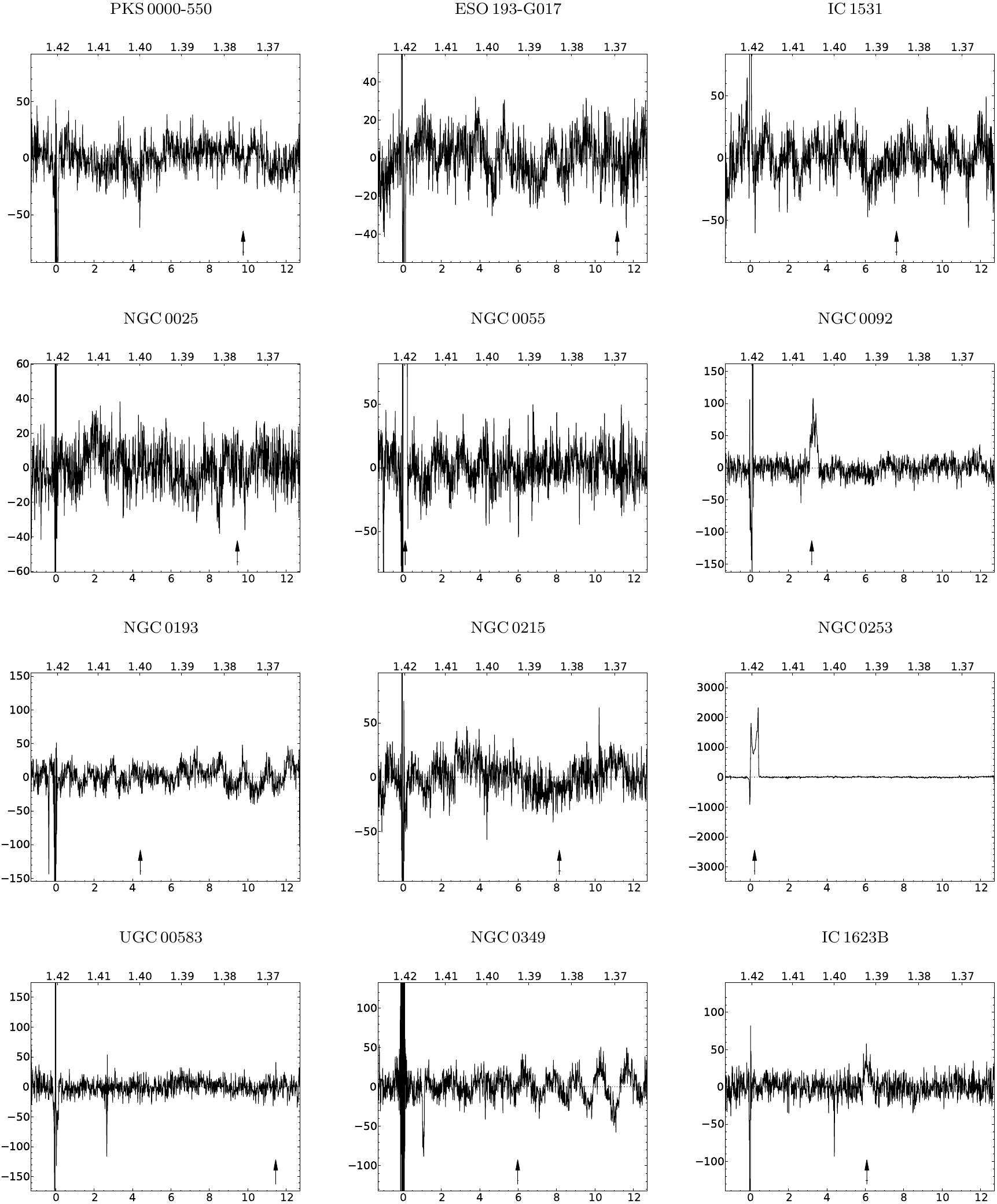}
\caption{HIPASS spectra of 204 galaxies searched for \mbox{H\,{\sc i}}
  absorption. For each spectrum, the ordinate shows the beam-weighted
  flux density (in mJy), the lower abscissa the Doppler corrected
  barycentric redshift (in 1000\,km\,s$^{-1}$) and the upper abscissa
  the observed frequency (in GHz). The arrow and horizontal line
  indicate the mean and uncertainty in the systemic redshift of the
  galaxy. The full version of this figure is available
  online.}\label{figure:spectra_total}
\end{figure*}

\twocolumn

\bsp

\label{lastpage}

\end{document}